\documentclass[10pt]{article}
\usepackage[sc]{mathpazo} %Like Palatino with extensive math support
\usepackage{fullpage}
\usepackage[authoryear,sectionbib,sort]{natbib}
\usepackage[utf8]{inputenc}
\usepackage{color}
\usepackage{amsmath}
\usepackage{graphicx}
\usepackage{float}

\title{Demographic stochasticity and resource autocorrelation \mbox{control} biological invasions in heterogeneous landscapes}

\author{Andrea Giometto$^{1,2,\ast}$ \\ 
Florian Altermatt$^{2,3,\ast}$ \\ 
Andrea Rinaldo$^{1,4}$}

\date{}

\begin{document}

\maketitle

\noindent{}1. Laboratory of Ecohydrology, School of Architecture, Civil and Environmental Engineering, \'Ecole Polytechnique F\'ed\'erale de Lausanne, CH-1015 Lausanne, Switzerland;

\noindent{}2. Eawag, Swiss Federal Institute of Aquatic Science and Technology, Department of Aquatic Ecology, CH-8600 D\"ubendorf, Switzerland;

\noindent{}3. Department of Evolutionary Biology and Environmental Studies, University of Z\"urich, CH-8057 Z\"urich, Switzerland;

\noindent{}4. Dipartimento di Ingegneria Civile, Edile ed Ambientale, Universit\`a di Padova, I-35131 Padua, Italy. \looseness=-1

\noindent{}$\ast$ Corresponding author; e-mail: andrea.giometto@epfl.ch and florian.altermatt@eawag.ch.

\bigskip

\noindent \textit{Keywords}: Environmental stochasticity, Biological invasions, Traveling waves, Front propagation, Fisher-Kolmogorov, Dispersal.

\newpage{}

\section*{Abstract}
Classical models of biological invasions assess species spread in homogeneous landscapes by assuming constant growth rates and random local movement. Mounting evidence suggests, however, that demographic stochasticity, environmental heterogeneity and non-random movement of individuals affect considerably the spread dynamics. Here, we show that the dynamics of biological invasions are controlled by the spatial heterogeneity of the resource distribution. We show theoretically that increasing the landscape resource autocorrelation length causes a reduction in the average speed of species spread. Demographic stochasticity plays a key role in the slowdown, which is streghtened when individuals can actively move towards resources. The reduction in the front propagation speed is verified in laboratory microcosm experiments with the flagellated protist \textit{Euglena gracilis} by comparing spread in habitats characterized by different resource heterogeneity. Our theoretical and experimental findings highlight the need to account for the intrinsic stochasticity of population dynamics to describe spread in spatially extended landscapes, which are inevitably characterized by heterogeneous spatial distributions of resources controlling vital rates. Our work identifies the resource autocorrelation length as a key modulator and a simple measure of landscape susceptibility to biological invasions, with implications for predicting the characters of biological invasions within naturally heterogeneous environmental corridors.

\newpage{}

\section*{Introduction}

Environmental fluctuations and heterogeneity are ubiquitous in nature and are thought to affect nearly all aspects of ecology, ranging from species coexistence to population synchrony, driving range shifts and potentially causing abrupt biotic change (e.g., \citealt{with95,with02,nelson12}). Local population dynamics in fluctuating and heterogeneous environments have been studied extensively in recent years (\citealt{gonzalez02,duncan13}), mainly with respect to population synchrony (\citealt{benton01,vasseur09,fox11}). Both theoretical (\citealt{roy05,vasseur07}) and experimental (\citealt{gonzalez02,fontaine05,massie15}) studies have highlighted the relevance of the temporal autocorrelation structure of environmental fluctuations for ecological dynamics. The study of ecological processes in the presence of environmental stochasticity at different levels of autocorrelation is of interest not only because environmental fluctuations are typically positively correlated (\citealt{beninca11}), but also in view of the global shift towards `bluer' climate variables (i.e., more fluctuating) across most continents (e.g., \citealt{garcia-carreras11}). Whereas most experimental investigations focused on temporal environmental fluctuations, spatial heterogeneity received surprisingly little attention (\citealt{with02,melbourne07}). Accordingly, the study of the implications of environmental fluctuations for spatial dynamics (\citealt{gonzalez02,duncan13}) and especially for the propagation of biological invasions (\citealt{neubert00,with02,mendez11}) is a challenging avenue for experimental research.

The effect of environmental fluctuations and spatial heterogeneity may be especially relevant in the context of biological invasions and range shifts, which are seen as some of the most relevant current dynamics across all ecosystems (\citealt{hastings05}). The spatial spread of invasions has been investigated extensively in the literature, starting with the pioneering works of Fisher, Kolmogorov and Skellam  (\citealt{fisher37,kolmogorov37,skellam51}). Traditionally, the propagation of invasive fronts has been modeled with the Fisher-Kolmogorov equation (\citealt{fisher37,kolmogorov37}) that predicts a linear rate of spread in homogeneous environments. Such equation was applied extensively to describe field data  (\citealt{lubina88,andow90}) and found applications also in other disciplines, for example physics and chemistry  (\citealt{mendez10}). Comprehensive reviews of mathematical modeling and empirical studies of species spread exist  (\citealt{hastings05}). Stochastic generalizations of the Fisher-Kolmogorov equation showed that demographic stochasticity affects the propagation dynamics, causing a reduction in the front propagation speed  (\citealt{hallatschek09}). Other modeling approaches have shown that temporal fluctuations in mean dispersal distances can increase invasion speed, while temporally uncorrelated fluctuations in demographic parameters typically decrease the front propagation velocity  (\citealt{ellner12,mendez11}). Despite the fact that most natural environments are inevitably heterogeneous (e.g., \citealt{holyoak05b}), however, much of the current understanding of species spread is based on theoretical models (\citealt{hastings05,mendez10}) that considered homogeneous landscapes. Only in recent years, progress has been achieved in the theoretical understanding of species spread in more complex, heterogeneous or fluctuating landscapes (\citealt{nelson98,neubert00,melbourne07,dewhirst09,mendez10,pachepsky11}), and such progress calls for experimental verification (\citealt{seymour14}). For example, thresholds for the minimal percentage of favorable habitat that can support spread have been studied (\citealt{with95,with02,dewhirst09}). Dewhirst and Lutscher (2009), for example, derived quantitative relationships for the invasion threshold and spread rates in integro-differential equation models in fragmented landscapes. The speed of biological invasions has been claimed to be affected by environmental stochasticity (\citealt{mendez11}) and an extensive line of research addressed the contribution of geometrical heterogeneities of the landscape to the propagation of invading fronts (\citealt{mendez03,mendez04c,campos06,bertuzzo07,seymour14}) suggesting that, in general, geometrical heterogeneities slow the speed of front propagation.

Integrating environmental heterogeneity in models of spread is a challenging task and a modeling framework that allows drawing general conclusions is lacking to date (\citealt{hastings05,urban08}). In the search for such a framework, the study of biological invasions in heterogeneous and fluctuating environments has been addressed in the context of the Fisher-Kolmogorov equation (\citealt{fisher37,kolmogorov37}) either by embedding various sources of environmental stochasticity in the original deterministic equation (\citealt{shigesada86,mendez03,mendez11}) or by considering spread in spatially heterogeneous media (\citealt{mendez03,mendez04c,campos06,bertuzzo07}). Environmental stochasticity and spatial heterogeneity (\citealt{nelson98,nelson12}) have been incorporated in the Fisher-Kolmogorov equation through noise terms that were uncorrelated in space, periodic in space (\citealt{shigesada86,kinezaki03}) or else characterized by a gaussian spatial correlation function with a fixed correlation length (\citealt{mendez11}). Whereas the importance of the autocorrelation structure of temporal environmental fluctuations for local ecological processes is now widely recognized (\citealt{gonzalez02,fontaine05,vasseur07,garcia-carreras11}), the effect of the spatial autocorrelation of environmental fluctuations on biological spread rates has just begun to be explored (\citealt{urban08}). The experimental study of species spread has recently started to test theoretical predictions of the Fisher-Kolmogorov model in homogeneous habitats (\citealt{croze11,korolev11,simpson13,giometto14}). A limited number of empirical works has measured spread rates in heterogeneous and diverse habitats and compared realized spread distances in patchily distributed sites (\citealt{bergelson94,bailey00,williamson02}). However, the results of these studies were not linked to Fisher-Kolmogorov-like models embedding environmental stochasticity or heterogeneity. In particular, experimental studies investigating the role of the resource autocorrelation structure in driving the spread of species are lacking.

Here, we study biological invasions in the presence of spatially heterogeneous resource distributions, which could, for example, reflect the spatial composition and quality of soil or topographically determined habitat elements such as exposure or elevation, or habitat fragmentation due to human land-use (e.g., \citealt{with95,with02}). Motivated by previous research on environmental fluctuations mentioned above, we focus on the effect of the spatial autocorrelation structure of the resource distribution on the propagation speed of biological invasion fronts. The distribution of resources is assumed to affect both the growth dynamics and movement behavior of individuals. Giometto et al. (2014) showed that, in homogeneous landscapes, demographic stochasticity introduces a noise term in the reaction-diffusion equation describing the front propagation, leading to a quantifiable variability of the process across replicated experimental invasions. Therefore, our tenet is that both environmental and demographic stochasticity jointly affect biological invasions and thus the interplay between these two sources of stochasticity is specifically investigated here.

We first show theoretically that the speed of species spread decreases when the resource autocorrelation length increases, all other conditions being equal. Second, we verify such prediction in a microcosm experiment with the flagellated protist \textit{Euglena gracilis}, by manipulating light intensity profiles along linear landscapes (light is an energy source for \textit{E. gracilis}, as it has chloroplasts and can photosynthesize). Third, we discuss the contribution of each process included in the model to the propagation of biological invasions. We show theoretically that demographic stochasticity is necessary to produce the slowdown, which is more pronounced if individuals can direct their movement towards resources. 

\section*{Methods}

\subsection*{Model}
Species spread in heterogeneous linear landscapes is modeled via a stochastic generalization of the Fisher-Kolmogorov equation including demographic stochasticity (\citealt{dornic05,bonachela12,giometto14}):
\begin{equation}
\frac{\partial \rho}{\partial t} = D \frac{\partial^2 \rho}{\partial x^2} + r(I) \rho \left[ 1-\frac{\rho}{K}\right] +\sigma \sqrt{\rho} \ \eta,
\label{FK_env1}
\end{equation}
where $\rho(x,t)$ is the density of individuals, $D$ is the diffusion coefficient of the species driven by the active movement of individuals, $r$ is the growth rate, $K$ is the carrying capacity, $\sigma$ is a parameter describing the amplitude of demographic stochasticity and $\eta$ is a gaussian, zero-mean white noise (i.e., the noise has correlations $\langle\eta(x,t)\eta(x',t')\rangle=\delta(x-x')\delta(t-t')$, where $\delta$ is the Dirac's delta function). It\^{o}'s stochastic calculus is adopted, as appropriate for the demographic noise term (\citealt{giometto14}). The growth rate $r(I)=r_0 I$ is assumed to be a function of the local amount of resources $I(x)$, which can assume two values: $I(x)=1$ or $I(x)=0$. Landscape heterogeneity is thus embedded in the resource profile $I(x)$. We studied the dimensionless form of equation (\ref{FK_env1}), which reads (see app. A available online):
\begin{equation}
\frac{\partial \rho'}{\partial t'} = \frac{\partial^2 \rho'}{\partial x'^2} + \chi_I \rho' \left[ 1-\rho' \right] +\sigma'' \sqrt{\rho'} \ \eta,
\label{adimenseq}
\end{equation}
where $t' = r_0t$, $x' = \sqrt{\frac{D}{r_0}}x$, $\rho'=\rho/K$, $\sigma''=\frac{\sigma}{\sqrt{K}(rD)^{1/4}}$ and $\chi_I(x')$ is the indicator function of the set of $x'$ for which $I(x')=1$. In the following we drop primes for convenience: one can recover the original dimensions by multiplying $t$ by $r_0$, $x$ by $\sqrt{r_0/D}$ and rescaling $\rho$ and $\sigma$ as indicated above. Numerical integration of stochastic partial differential equations with square root noise terms require ad hoc numerical methods, as standard approaches such as the first-order explicit Euler method inevitably produce unphysical negative values for the density $\rho$  (\citealt{dornic05}). Therefore, equation (\ref{adimenseq}) was integrated with the split-step method proposed in Dornic et al. (2005), see app. A for details.

We generated landscapes with various resource autocorrelation lengths by imposing $I(x)$ to be composed of subsequent independent patches of suitable ($I(x)=1$ and $r=r_0$) or unsuitable ($I(x)=0$ and $r=0$) habitats (fig. \ref{model}\textit A). The length of each patch was drawn from an exponential distribution with rate $\mu$. Therefore, each landscape was a stochastic realization of the so-called telegraph process with rate $\mu$ and autocorrelation length $c_L=1/(2\mu)$. The mean extent of suitable and unsuitable patches in such landscapes is $1/\mu$. Because simulated landscapes were finite, we only accepted landscapes with mean resources equal to $\bar I = L^{-1} \int_0^L  I(x) dx= 1/2$ and autocorrelation length confined to a narrow window around $1/(2\mu)$. Examples of landscapes used in the simulations are shown in fig. \ref{model}\textit A.

We generated $96$ landscapes for each value of resource autocorrelation length $c_L$ and integrated equation (\ref{adimenseq}) numerically for each landscape and for each value of $\sigma\in\{0.1,0.2,0.4,0.6\}$ (fig. \ref{model}\textit B), with initial density profiles localized at the origin. To avoid the extinction of the whole population, we fixed the left boundary at $\rho=1$. For each numerical integration, we measured the position of the front by fixing a threshold value of the density ($\bar \rho = 0.15$) and recording the furthest point from the origin where the cell density was higher than such value. The mean propagation speed for each value of the resource autocorrelation length was computed by fitting a straight line (least-squares fit) to the mean front position versus time in the asymptotic propagation regime (fig. \ref{mean_front}), before any of the replicated invasions reached the end of the landscape.

We derived a theoretical approximation to the mean front propagation speed, valid for large autocorrelation lengths $c_L$ and $\sigma$, by characterizing the mean time taken to cross a patch of unfavorable habitat (where $I=r=0$) of length $z$. Such mean time is shown (app. A) to depend on $z$ and $\sigma$ as $\langle \tau \rangle(z,\sigma)=C z^2 e^{ d\left( z \sigma^b \right)^a}$,
where $C$, $a$, $b$ and $d$ are constants, independent of $z$ and $\sigma$. Additionally, we characterized the functional dependence of the variance of $\tau$ on $z$ and $\sigma$ and derived an approximation to the variance of the total time taken by a front to colonize completely a landscape of finite length $L$ (app. A). Our approximation is in good agreement with numerical integrations of equation (\ref{adimenseq}) (fig. \ref{variance_sq_exp}).

To test whether deterministic models predict a slowdown of the invading front for increasing resource autocorrelation length, we numerically integrated equation (\ref{adimenseq}) with $\sigma=0$. Additionally, we numerically integrated equation (\ref{adimenseq}) with $\sigma=0$ and imposing a negative growth rate $r$ in unfavorable patches where $I=0$ (app. A).

\subsection*{Experiments}
We performed experiments with the flagellated protist \textit{Euglena gracilis}, acquired from Carolina Biological Supply (NC, USA). A culture of \textit{E. gracilis} was initialized two weeks prior to the start of the experiment and kept at $22$ $^\circ$C under constant LED (Light Emitting Diode, model SMD 5050) light of wavelength $469$ nm (emission width approximately $10$ nm), in a filtered ($0.2$ $\mu$m filter) nutrient medium composed of sterilized spring water and Protozoan Pellets (Carolina Biological Supply, NC, USA) at a density of $0.45$ g$\cdot$l$^{-1}$ in a $500$ ml Schott flask (\citealt{altermatt15}).

In our experiment, light was used as the energy source for \textit{E. gracilis}. To demonstrate that light was crucial for the growth of \textit{E. gracilis} in our experimental setting, we measured \textit{E. gracilis}' growth curves (fig. \ref{experiment}\textit A) by initializing eight low-density cultures in $10$ ml cell culture flasks. Half of such cultures were placed on top of two LEDs (for each culture) operated at a total flux of $1$ mW each. The other half of the cultures were placed on top of two LEDs (for each culture) operated at the same power, but covered with black tape so that no light would penetrate. The spatial arrangement of illuminated and non-illuminated cell culture flasks was randomized.

Light also affects the movement behavior of \textit{E. gracilis} individuals through a process known as phototaxis, the directed movement of cells towards or away from light (\citealt{drescher10,giometto15}). Specifically, at low to intermediate light intensities, \textit{E. gracilis} swims towards the light source at a time scale much shorter than the typical generation time. At very high light intensities, negative phototaxis can also be observed, and the plastic reaction of phototaxis can be induced very reliably (\citealt{giometto15}). The light intensity value used in our experiments is smaller than the light intensity value at which negative phototaxis occurs.

The front propagation experiment was performed in linear landscapes, which were channels drilled on a plexiglass sheet ($5$ mm wide, $3$ mm deep and $1.9$ m long, respectively, $300$, $200$ and $10^5$ times the size of an individual, see \citealt{giometto13}), filled with filtered nutrient medium (fig. \ref{setup}\textit A). A gasket avoided water spillage and a plexiglass lid was used to seal the system. The experimental replicates were kept in a climatized room at $22$ $^\circ$C for the whole duration of the experiment. Heterogeneous distributions of resources were generated via linear arrays of LEDs (fig. \ref{setup}\textit B) controlled via Arduino Uno boards. LEDs in the array were separated by a distance of $\Delta L=3.12$ cm from each other and could be switched on or off individually. Switched-on LEDs emitted light with an intensity of $5.2$ W$\cdot$m$^{-2}$ within the plexiglass channel, immediately above the LED. The linear landscapes were placed on top of the LED array at a distance of $4.5$ mm. The light intensity profile generated by one LED was measured by placing a white paper sheet inside the plexiglass channel and by measuring the irradiance on the sheet with a digital camera operated in grayscale. The total radiant flux of the LED was measured via a calibrated photodiode. Light intensity profiles with the desired autocorrelation length were designed by imposing the probability $\lambda$ of the LED number $i+1$ in the LED array to be switched-on if the LED number $i$ was switched-off, that is, $\mathbb{P}[$LED$(i+1)=$ON $|$ LED$(i)=$OFF$]=\lambda$. Such Markov Chain was imposed to be symmetric, that is, $\mathbb{P}[$LED$(i+1)=$OFF $|$ LED$(i)=$ON$]=\lambda$. Small and large values of $\lambda$ generate resource distributions with long and small autocorrelation lengths (approximately equal to $\Delta L/(2 \lambda)$), respectively. Because landscapes were of finite total length, the above procedure could generate by chance resource profiles with autocorrelation length different from the desired one and with a mean frequency of switched-on LEDs different from $1/2$. Therefore, the set of resource profiles obtained with the above Markov Chain procedure was restricted to those with a mean frequency of switched-on LEDs equal to $1/2$ and in a narrow window of autocorrelation length around the desired one. Therefore, all replicates had the same mean light intensity $\bar I(x) = L^{-1} \int_0^L I(x)dx$.

We compared two treatments in the experiment. Treatment \textit{1} consisted of landscapes with identical small autocorrelation length ($c_L \simeq 2$ cm) but different switched-on LED sequences, generated via the Markov Chain procedure with $\lambda=0.75$. Treatment \textit{2} consisted of landscapes with identical large autocorrelation length ($c_L\simeq 6$ cm) but different switched-on LED sequences, generated via the Markov Chain procedure with $\lambda=0.25$. The choice of the large autocorrelation length value in the experiment was limited by the total finite length of the experimental setup and was chosen to be less than $1/20$ of the total setup length. We initially had six landscape replicates of each treatment, but lost one replicate of Treatment \textit{1} due to leakage. All $11$ landscapes had the same total number of switched-on LEDs and the experimental light intensity profiles are shown in fig. \ref{experiment_light}. The stated values of autocorrelation length are based on the first-order autocorrelation of the Markov Chain that generated the landscape. The first three LEDs in every landscape were switched-on to allow the local establishment of the inoculated \textit{E. gracilis} population and to avoid differences between the two treatments in the initial establishment dynamics. Thus, the landscapes generated via the Markov Chain procedure described in the text started at the fourth LED. In Treatment \textit{2} (large autocorrelation length), three landscapes were chosen so that the fourth LED was switched on and the other four were chosen so that the fourth LED was switched off. In other words, the realized Markov Chain started from its stationary distribution. The spatial arrangement of landscapes belonging to the two treatments on the experimental bench was randomized.

At the start of the experiment, we introduced an ensemble of \textit{E. gracilis} individuals at one end of the linear landscapes. Following the inoculation, we measured for eight consecutive days the density of \textit{E. gracilis} throughout all replicates by taking pictures with a stereomicroscope (model Olympus SZX16 with the digital camera Olympus DC72) and counting individuals via image analysis (\citealt{altermatt15}).

\subsection*{Statistical analysis}
We used a mixed effect model to compare the speed of the propagating \textit{E. gracilis} among the two different treatments. Thereby, the autocorrelation treatment was included as a fixed effect, while day and replicate were included as random effect. We repeated this analysis using different choices of threshold values used for determining the front position. The minimum and maximum threshold values employed in the statistical analysis were chosen such that no replicate displayed a retreating front between successive measurements (caused by noise in the density profiles). The test statistics are reported in table \ref{me_stats} for the density threshold value $\bar \rho = 60$ cm$^{-1}$ and in table \ref{me_stats_all} for all values of $\bar \rho$ considered. We did not include the first timepoint in the analysis because it was measured immediately after the inoculation of \textit{E. gracilis} in the landscape and thus was identical for all replicates. Because the propagating front reached the end of the landscape at day $4$ in some replicates, the front propagation analysis was performed only with the data up to day $3$ (included) to avoid spurious border effects due to the finite size of the system.

\subsection*{Model with directed movement towards resources}

Equation (\ref{FK_env1}) does not assume directed movement of individuals towards resources; such directed movement, however, occurs in our experiment and is likely to occur in nature (\citealt{andow90,fronhofer13}). Additionally, the experimental resource distributions (i.e., the light intensity profiles $I(x)$) were not simply sequences of illuminated and non-illuminated spatial patches with sharp edges, but, rather, smooth light intensity profiles alternating between well-lit and dark regions of the landscape according to the spatial arrangement outlined above. Because \textit{E. gracilis} is capable to detect light intensity gradients and to move towards well-lit regions of the landscape, such directed movement may affect the invasion dynamics. To assess the net contribution of the directed movement of individuals towards resources, we incorporated in equation (\ref{FK_env1}) the model for phototaxis derived in Giometto et al. (2015). The phototactic term was inferred from measurements of stationary density distributions of \textit{E. gracilis} in the presence of light gradients and was shown to reproduce the accumulation dynamics of \textit{E. gracilis} populations accurately in Giometto et al. (2015). The model equation reads:
\begin{equation}
\frac{\partial \rho}{\partial t} = \frac{\partial}{\partial x} \left[ D \frac{\partial \rho}{\partial x} - \frac{d \phi}{dx}(I) \rho \right] + r(I) \rho \left[ 1-\frac{\rho}{K}\right] +\sigma \sqrt{\rho} \ \eta,
\label{FK_env}
\end{equation}
where $\phi=a(I-I_c)/(I+I_r)$ is the phototactic potential describing \textit{E. gracilis}' attraction towards (or against) light (\citealt{giometto15}). The parameters describing $\phi$ were estimated (\citealt{giometto15}) and were set equal to $a=1.4 \cdot 10^{-8}$ m$^4\cdot$W$^{-1}\cdot$s$^{-1}$, $I_r=1.7$ W$\cdot$m$^{-2}$ and $I_c=28$ W$\cdot$m$^{-2}$. 
We assumed that $r$ follows Monod kinetics (the assumption is customary for phytoplankton, \citealt{diehl02}), that is, $r(I)=r_1 I/(I+K_I)$, where $K_I$ is the half-saturation constant. The model (equation \ref{FK_env}) was integrated with parameters suitable to describe the experimental system, $r_1=6 \cdot 10^{-3}$ min$^{-1}$, $K_I=1$ W$\cdot$m$^{-2}$,  $K=300$ cm$^{-1}$, $D=0.08$ cm$^2\cdot$min$^{-1}$ (estimated in  \citealt{giometto15}), various values of $\sigma$ (fig. \ref{speed_photo}) and initial condition localized at the origin. See app. A for  details on the numerical integration scheme adopted. The slowdown effect caused by the resource autocorrelation structure is also found with other choices of the growth rate dependence on the resource density. In fact, we found that results do not change qualitatively by assuming a linear dependence of $r$ on $I$. We used equation (\ref{FK_env}) to simulate biological invasions in linear landscapes with resource distributions $I(x)$ exhibiting various autocorrelation lengths. To mimic the experimental setup (fig. \ref{experiment_light}), such landscapes were generated with the same Markov-chain procedure used to design the experimental landscapes (see \textit{Experiment} section), where the light intensity profile generated by a single LED (centered in $x=0$) was assumed equal to the best fit of the equation $I(x)=c_0 / (c_1^2+x^2)^2$ to the measured light intensity profile (see fig. S1 of \citealt{giometto15}). The total light intensity was kept constant for all landscapes. To further mimic the experiment, we set reflecting boundary conditions for the integration of equation (\ref{FK_env}) and simulations in which the population went extinct were excluded from the analysis. Therefore, the model equation (\ref{FK_env}) was specifically derived to reproduce as closely as possible the experimental system at hand. Landscapes used in the simulations were much longer ($18$ m) than those used in the experiment in order to avoid border effects. Such numerical settings allowed a clear identification of the invasion front and allowed simulating species spread in landscapes with very large autocorrelation length, which could not be investigated experimentally because of the finite size of the experimental setup. 

\section*{Results}

\begin{figure}[H]
\centering
\includegraphics{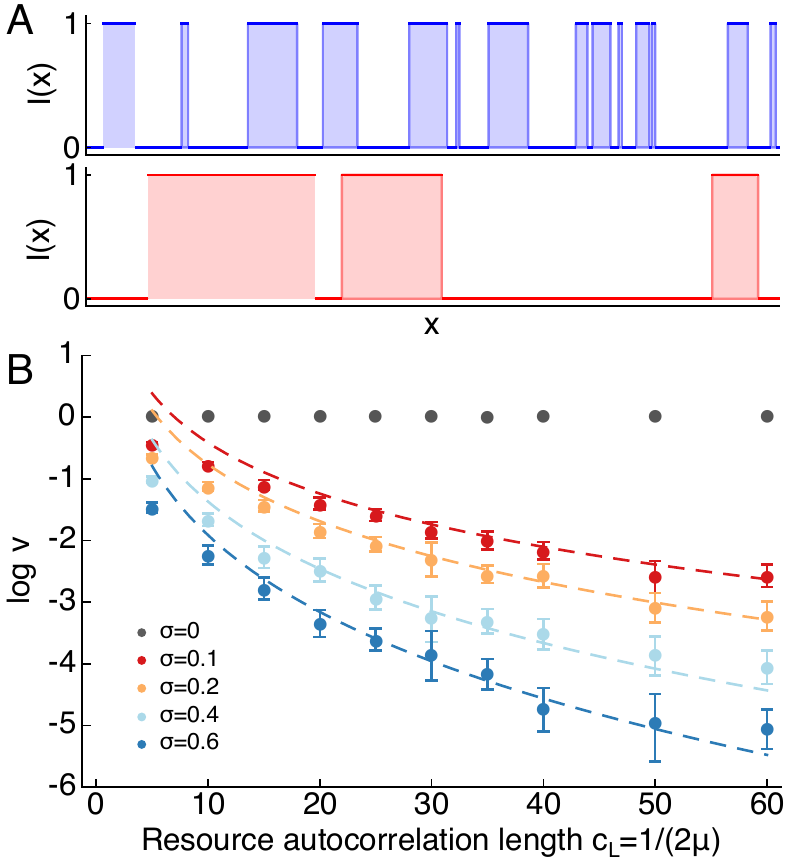}
\caption{Mean front propagation in the model (dimensionless equation \ref{adimenseq}). (\textit A) Examples of landscapes with different resource autocorrelation length $c_L$, generated via the telegraph process with rate $\mu$ (Methods). (\textit B) The mean invasion speed computed in numerical integrations of the model (equation \ref{adimenseq}) decreases with increasing resource autocorrelation length $c_L$ (log-linear plot) for $\sigma>0$ and is a decreasing function of the amplitude of demographic stochasticity $\sigma$ (different colors according to legend). With $\sigma=0$ the dynamics is deterministic and the mean front propagation speed does not decrease with $z$ (gray dots). Error bars display the $95$\% confidence interval for $\log v$, computed with $2 \cdot 10^3$ bootstrap samples. Error bars for $\sigma=0$ are smaller than symbols. Dashed lines show the mean front propagation speed computed according to the theoretical approximation (equation \ref{speed}).}
\label{model}
\end{figure}

\begin{figure}[H]
\centering
\includegraphics[width=11cm]{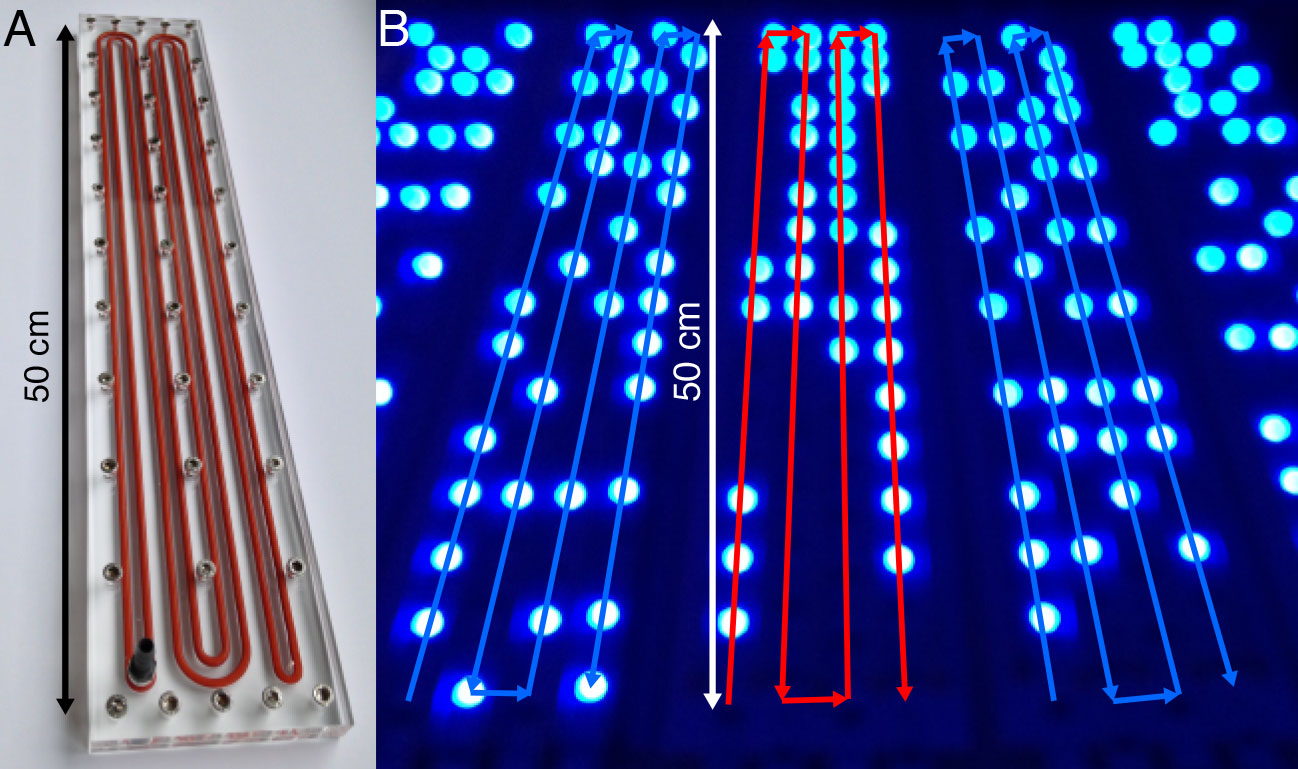}
\caption{Experimental setup. (\textit A) Linear landscapes used in the experiments were channels drilled on a plexiglass sheet. A gasket (orange rubber band) avoided water spillage. (\textit B) Photograph of the LED strips used to control the distribution of resources for \textit{E. gracilis}. The red and blue lines show the paths of landscapes with large and small resource autocorrelation length, respectively.}
\label{setup}
\end{figure}

\begin{figure}[H]
\centering
\includegraphics{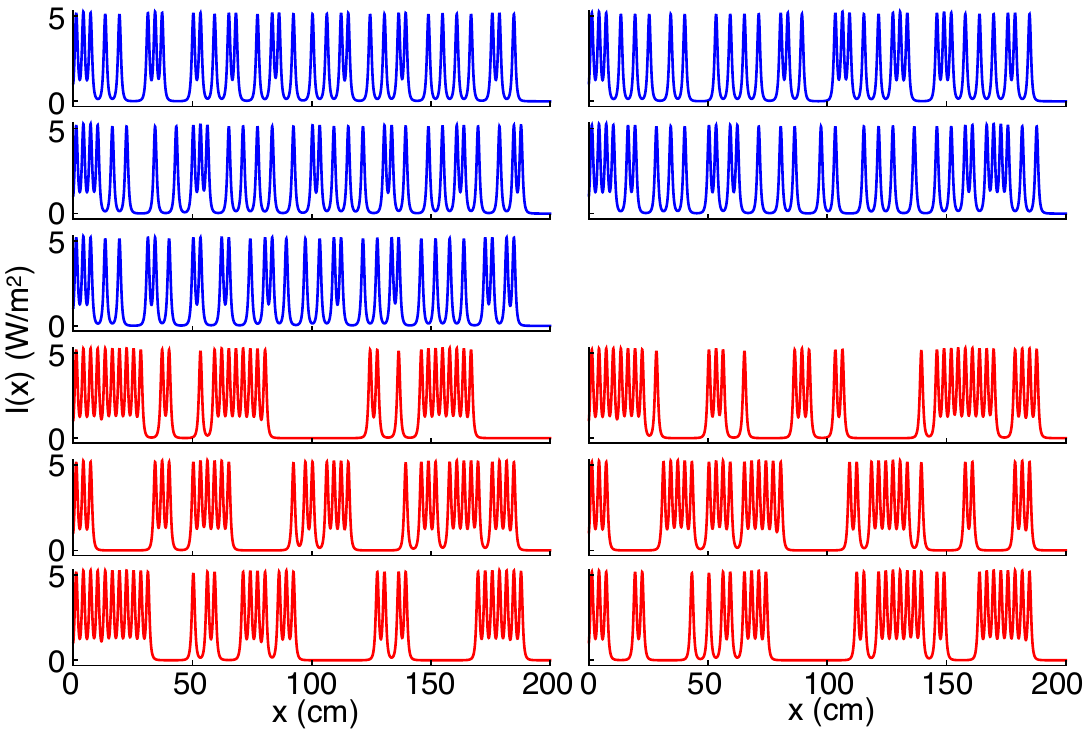} 
\caption{Light intensity profiles used in the experiment (Methods). One spread experiment was performed for each landscape. The total light intensity is the same for each landscape. Landscapes with the same color have identical small (blue) or large (red) autocorrelation length of the resource distribution $I(x)$, but different LED on-off sequences.}
\label{experiment_light}
\end{figure}

Our generalization of the Fisher-Kolmogorov equation (equations \ref{FK_env1} and \ref{adimenseq}) includes demographic stochasticity and resource heterogeneity. Such resource heterogeneity affects the spread dynamics through the dependence of the growth rate $r$ on the local amount of resources $I$ (Methods). We found that the speed of invasion in the model equation (\ref{adimenseq}) decreases with increasing resource autocorrelation length (fig. \ref{model}\textit B). The mean front propagation speed, in heterogeneous landscapes where resource patch lengths are distributed exponentially with rate $\mu$, depends on $c_L$ and $\sigma$ asymptotically (i.e., for large $c_L$ and $\sigma$) as:
\begin{equation}
v=\frac{L}{ \frac{\mu L}{2} \int_0^L dz \langle \tau \rangle(z,\sigma) \mu e^{-\mu z} } \simeq \frac{8 c_L^2} {{ \int_0^\infty dz \langle \tau \rangle(z,\sigma) e^{-z/(2c_L)}}}.
\label{speed}
\end{equation}
Figs. \ref{model}\textit B and \ref{speed_fig} show that equation (\ref{speed}) correctly predicts the speed of invasion at large values of $c_L$ and $\sigma$. In heterogeneous landscapes with different spatial arrangements of favorable and unfavorable patches, if the percentage of space occupied by unfavorable patches is $f_0 \in (0,1)$ and the distribution of such patches lengths is $p_0(z)$, with mean $\int dz z p_0(z)=1/\mu$, the asymptotic invasion velocity can be approximated as:
\begin{equation}
v=\frac{1}{{\mu f_0 \int_0^\infty dz \langle \tau \rangle(z,\sigma) p(z)}}.
\label{general_speed}
\end{equation}
We show in the app. A that equation (\ref{general_speed}) correctly predicts the speed of invasion in landscapes with percentages of unfavorable habitat different from $f_0=1/2$ (fig. \ref{mean1323}). Note that the speed of invasion according to equations (\ref{speed},\ref{general_speed}) is a function of the autocorrelation length if the landscapes consist of favorable and unfavorable patches generated through the telegraph process outlined in the Methods section. In general, however, the speed of invasion is not a univocal function of the resource autocorrelation length (or of other characteristic length scales of the landscape), but it rather depends on the whole distribution of unfavorable patch lengths through equation (\ref{general_speed}). The slowdown effect is due to the fact that, in the presence of demographic stochasticity, long patches of unfavorable habitat act as obstacles for the spread of populations. The larger the extent of the unfavorable patch, the longer it takes for a population to cross it. The front propagation speed is also found to be a monotonically decreasing function of the amplitude of demographic stochasticity (fig. \ref{model}\textit B). Accordingly, integrating the model without demographic stochasticity ($\sigma=0$ in equation \ref{adimenseq}, gray dots in fig. \ref{model}\textit B) leads to no discernible slowdown of the front in strongly autocorrelated versus weakly autocorrelated landscapes, even when imposing negative values of the growth rate $r$ in unfavorable patches where $I=0$ (app. A). Such results demonstrate that the local extinctions caused by demographic stochasticity in unfavorable patches are responsible for the observed front slowdown.

Numerical integration of equation (\ref{adimenseq}) shows that the variability of the front position increases for larger values of $c_L$ and $\sigma$. Such increased variability is caused by two factors: i) two landscapes with identical resource autocorrelation lengths appear increasingly dissimilar for increasing values of the typical patch length $1/\mu$; ii) the variance of the distribution of waiting times (i.e., the times to cross an unfavorable patch of length $z$) increases (approximately) quadratically with the mean time $\langle \tau \rangle(z,\sigma)$ (fig. \ref{stdtau}). These two observations can be used to approximate the fluctuations of the total time spent by the front to colonize a landscape of length $L$ (fig. \ref{variance_sq_exp}), as shown in the app. A.

The model (equations \ref{FK_env1} and \ref{adimenseq}) assumes random local movement of individuals. Although such assumption may be appropriate to describe spread in homogeneous landscapes (\citealt{andow90,giometto14}), individuals might be able to exploit local information on the availability of resources to direct their movement towards more favorable regions (\citealt{andow90,fronhofer13,fronhofer15}). We studied the effect of biased movement towards resources by including an advection term (towards regions endowed with more resources) in equation (\ref{FK_env1}), leading to equation (\ref{FK_env}). The latter model predicts again that the front propagation speed decreases for increasing resource autocorrelation length, in accordance with the former model (equation \ref{FK_env1}). Integrating equation (\ref{FK_env}) with and without the advection term shows that the biased local movement towards resources causes an increased slowdown of the invasion front in strongly (compared to weakly) autocorrelated landscapes (fig. \ref{speed_photo}). In other words, the biased movement towards resources acts as a spring that keeps the population in favorable patches and works against the exploration of unfavorable ones. Excluding demographic stochasticity from the model equation (\ref{FK_env}) leads again to the elimination of the slowdown effect (inset of fig. \ref{speed_photo}).

\begin{figure}[H]
\centering
\includegraphics{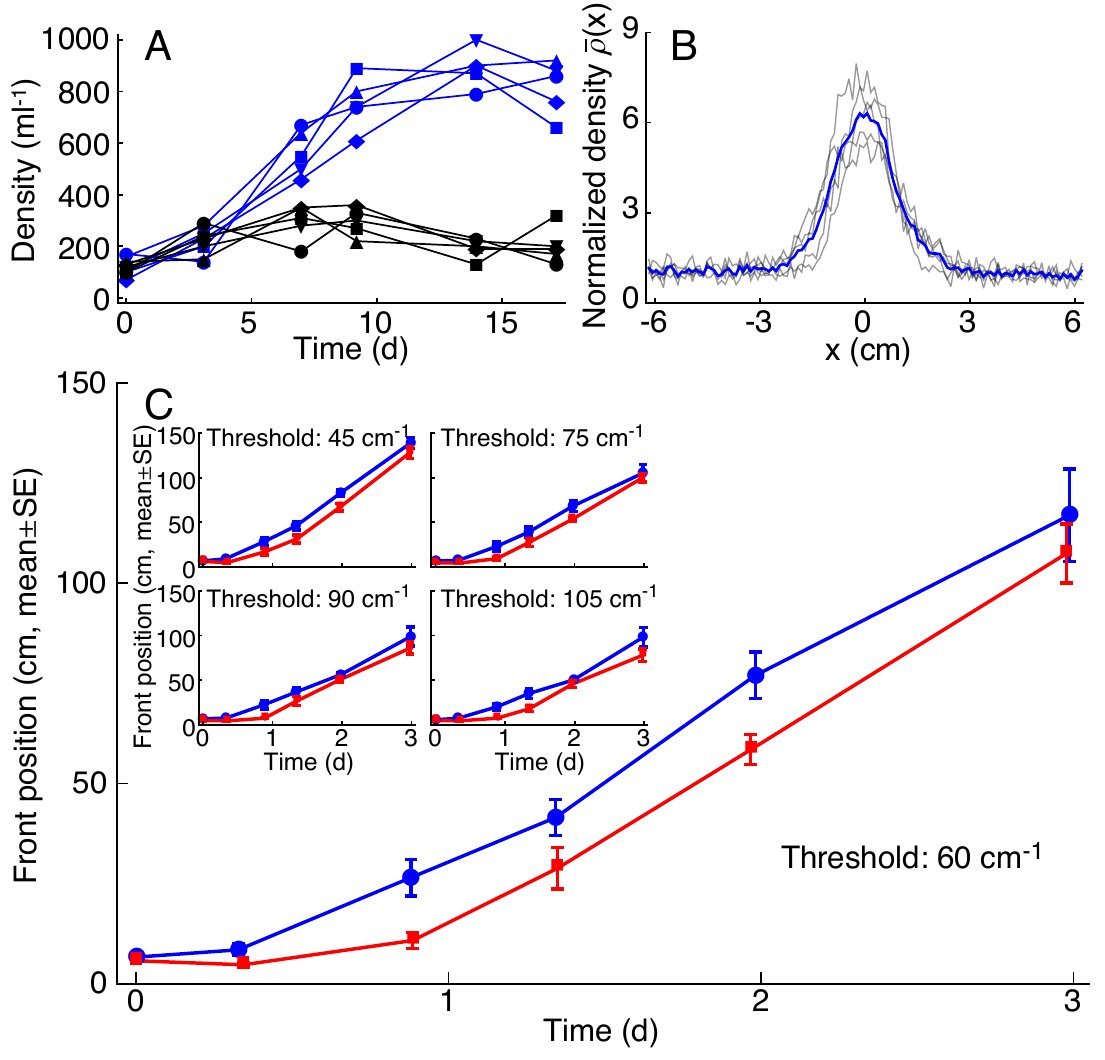}
\caption{Experimental spread in autocorrelated landscapes. (\textit A) Light was used as energy resource for \textit{E. gracilis}. Replicated measured growth curves show that \textit{E. gracilis} grows in the presence of light (blue symbols and lines) and does not grow in its absence (black symbols and lines). (\textit B) Replicated measurements (gray lines) of \textit{E. gracilis} density profiles (normalized by the value at the edge of the imaging window) in the presence of a LED at $x=0$ cm show that \textit{E. gracilis} populations accumulate around light sources through phototaxis. The blue line denotes the mean density profile across replicates (panel \textit B is redrawn from \citealt{giometto15}). (\textit C) Mean ($\pm$SE) position of the front, calculated among replicates with identical large (red) or small (blue) resource autocorrelation length at the threshold density value $\bar \rho = 60$ cm$^{-1}$. The inset shows mean front positions calculated at different threshold density values $\bar \rho$ as indicated. The slowdown effect is significant with all choices of $\bar \rho$, see Table \ref{me_stats_all}.}
\label{experiment}
\end{figure}

We designed an experiment with \textit{E. gracilis} to test the slowdown effect on the front propagation caused by the spatial resource autocorrelation length. We observed a steady front propagation across all landscapes with a mean front propagation speed of $54\pm9$ cm$/$d (mean$\pm$SE). The mean total number of individuals was $2420\pm110$ (mean$\pm$SE) at the start of the experiment (day $0$), $15000\pm800$ (mean$\pm$SE) at the end of the front propagation phase (day $4$) and $27000\pm4500$ (mean$\pm$SE) at the end of the experiment (day $8$). Thus, the invasion process was a combination of active, directed movement of individuals as well as reproduction. We found a significantly slower front propagation in landscapes in which the resources were strongly spatially autocorrelated (mixed effect model $p=0.027$, see also table \ref{me_stats}). The result is robust to changes of the threshold value at which the front position is evaluated (table \ref{me_stats_all}, figs. \ref{experiment}\textit{C} and \ref{experiment_all}). The slowdown effect is visible in fig. \ref{experiment}\textit C, which shows the mean front position across replicated invasions in the two treatments.

\section*{Discussion}

Our experiments show that the slowdown effect predicted by the stochastic models equations (\ref{FK_env1}), (\ref{adimenseq}) and (\ref{FK_env}) is found in microcosm experimental systems, which can be used to bridge theoretical models and natural systems  (\citealt{benton07}). In these experiments the demographic and movement traits of the study species were fixed and dictated by the species. The accompanying models additionally allowed to single out the individual role and the mutual interconnections of all processes included in the equations to the propagation dynamics in landscapes with different resource autocorrelation lengths.

Our theoretical and experimental investigation advances our current understanding of the spread of invading organisms in heterogeneous landscapes by addressing the joint effect of spatial environmental autocorrelation and demographic stochasticity on the spread dynamics. As arguably all natural landscapes are characterized by heterogeneous distributions of resources and all populations are subject to demographic stochasticity, our model incorporates two key elements hitherto often overlooked in the modeling of biological spread. A major result of our work is that demographic stochasticity is a key factor in the slowdown of front propagation in heterogeneous landscapes. Such finding highlights the importance of including demographic stochasticity in theoretical models because of the many facets through which it affects species spread (\citealt{hallatschek09,giometto14}). The implications of the above results challenge the standard approach as stochastic effects are neglected by deterministic, Fisher-Kolmogorov-like models. Because the slowdown effect is only observed when demographic stochasticity is included in the models, our theoretical investigation suggests that the stochastic birth-and-death dynamics are the main drivers of the observed reduction in propagation speed, rather than the movement behavior of individuals in heterogeneous landscapes that has received so far most attention in the literature (\citealt{morales02,vandyck05}). Previous studies have investigated the minimum percentage of suitable habitat that allows invasions to spread (\citealt{with95,with02,dewhirst09}), suggesting that invasions cannot propagate in landscapes with mean resource density below a critical threshold. Our work shows, complementarily, that the spatial arrangement of resources affects species spread even if the total amount of available resources is kept constant. Thus, it is not only the mean resource density that matters for the front propagation dynamics, because the autocorrelation structure of landscape heterogeneity alone also affects species spread. Our investigation extends previous works that addressed the effect of temporal environmental fluctuations on species spread  (\citealt{mendez11,ellner12}) by showing that the autocorrelation length of the resource distribution should be added to the environmental factors that can slow species spread, along with temporal fluctuations of vital rates  (\citealt{neubert00,ellner12}), geometrical heterogeneities of the substrate  (\citealt{mendez03,mendez04c,bertuzzo07}) and demographic stochasticity (\citealt{hallatschek09}).

Our finding that larger autocorrelation lengths reduce the spread rate of invading species is compatible with the results of Bergelson et al. (1994), who performed a field experiment with the invading weed \textit{Senecio vulgaris} and found that the average spatial distance between two generations along linear transects increased when favorable patches were uniformly distributed in space (in the parlance of our work, the transect featured a small autocorrelation length), compared to transects with clumped patches (i.e., endowed with large autocorrelation length). Bailey et al. (2000) performed spread experiments with the fungal plant pathogen \textit{Rhizoctonia solani}. Such work provides a complementing view to our investigation by evaluating the effect of the inter-distance between favorable patches on the spread and identifying experimentally the existence of a percolation threshold at a critical level of inter-patch distance. In the framework addressed here,  the analog of such percolation threshold corresponds to an autocorrelation length much larger than the average distance traveled by the front during one generation. There exist considerable differences in the experimental setup and the study system between this investigation and those in Bergelson et al. (1994) and Bailey et al. (2000). Most importantly, biased active movement towards favorable patches was present in the experiment performed here and embedded in equation (\ref{FK_env}), while passive dispersal was implemented in Bergelson et al. (1994). Both Bailey et al. (2000) and Bergelson et al. (1994) differ from this study because the landscape and the distribution of resources herein are continuous, whereas they adopted discrete spatial distributions of favorable patches. Although such discrete distributions might provide a good approximation to some fragmented landscapes, continuous heterogeneous distributions may be equally likely to occur in nature. Compared to previous experimental efforts, we provide a general theoretical framework to interpret the dynamical processes underlying the realized invasions. The theoretical investigation of equations (\ref{FK_env1}), (\ref{adimenseq}) and (\ref{FK_env}), in fact, allowed isolating the net effect of each process embedded therein. Furthermore, the theoretical approximation to the mean speed of invasion in the model (equation \ref{adimenseq}) derived here allows to quantitatively predict the dependence of such mean speed on the resource autocorrelation length $c_L$, the strength of demographic stochasticity $\sigma$ and the other species traits.

Our results have important implications for species spread in natural environments, which are generally characterized by resources (seen as any field controlling vital rates, especially reproductive ones) being heterogeneously distributed. The typical autocorrelation length of the resource distribution can be inferred from environmental data (\citealt{turner05,urban08}) and can be used as a concise indicator for the propagation success of a species of interest.  Furthermore, the spatial availability of resources is often altered by human activities, reinforcing the fragmentation of landscapes. In fact, habitat fragmentation may decrease significantly the autocorrelation length of the landscape through the introduction of qualitatively different patches in the natural environment (\citealt{with02,holyoak05b}). Our results give quantitative grounds to field observations on the effect of environmental heterogeneity on species spread. For instance, Lubina and Levin (1988) observed pauses in the spread of the California sea otter (\textit{Enhydra lutris}) in the presence of habitat discontinuities. Such pauses and the corresponding piecewise-linear propagation of the front (see fig. 2 of \citealt{lubina88}) are also found in our model (fig. \ref{front_simulations}), which enables to relate the mean spatial extent of habitat discontinuities to the average speed of invasion through equations (\ref{speed}) and (\ref{general_speed}). An alternation between phases of halt and spread was also found in the range expansion of the cane toad (\textit{Chaunus marinus}) in Australia (fig. 2 of \citealt{urban08}). Urban et al. (2008) performed an in-depth analysis of the effect of environmental heterogeneity on the spread of the cane toad in the field and found a statistically significant effect of environmental heterogeneity and, most importantly, of the spatial autocorrelation of environmental variables on the realized patterns of invasion speed. They found such effect in nature in a realized (not replicable) invasion, and thus they could only correlate the realized spread dynamics and its reduction with the landscape autocorrelation. Here, we have given a mathematical framework and an experimental proof showing that the slowdown effect caused by the spatial autocorrelation structure of the landscape is not an artifact of the mathematical model.

\section*{Conclusion}

In conclusion, our work demonstrates the need to account for the intrinsic stochasticity of population dynamics to broaden our understanding of ecological processes occurring in spatially extended natural landscapes, which typically display various degrees of heterogeneity. Further work should be dedicated to the modeling and experimentation of species spread in temporally-varying landscapes and, possibly, spatially-heterogeneous landscapes that fluctuate in time. Drawing from the literature on population dynamics in temporally-fluctuating environments, understanding the causal link between the autocorrelation structure of fluctuations and the dynamics of species spread is a promising direction for future research in this area.

\section*{Acknowledgments}

\noindent We thank Enrico Bertuzzo, Francesco Carrara, Lorenzo Mari and Amos Maritan for many useful discussions. We gratefully acknowledge the support by Swiss Federal Institute of Aquatic Science and Technology (Eawag) discretionary funds and Swiss National Science Foundation Projects 200021\_157174 and PP00P3\_150698.

\newpage{}

\bibliographystyle{amnatnat}
\bibliography{biblio}

\newpage{}

\section*{Tables}
\renewcommand{\thetable}{\arabic{table}}
\setcounter{table}{0}

\begin{table}[h]
\caption{Mixed-effect test statistics}
\begin{center}
\begin{tabular}{l|r|r|r|r|r}
\multicolumn{2}{ r |}{Value}	&	Std. Error	&	df	&	$t$-value	&	$p$-value	\\ \hline
Intercept	&	$45.98$	&	$3.27$	&	$44$	&	$14.04$	&	$p<10^{-4}$	\\
Autocorrelation length	&	$-11.61	$&	$4.43$	&	$9$	&	$-2.62$	&	$0.0279$
\end{tabular}
\end{center}
\label{me_stats}
\bigskip{}
{\footnotesize Mixed-effect test statistics testing the speed of front propagation, with the autocorrelation length treatment as single fixed effect and time/replicate as random effect. The treatment with small autocorrelation length had $5$ replicates, the treatment with large autocorrelation length had $6$ replicates. The front position was measured at the density threshold value $\bar \rho = 60$ cm$^{-1}$.}
\end{table}

\newpage{}

\section*{Figures}

%
%\subsection*{Online figure legends}
%
\renewcommand{\thefigure}{A\arabic{figure}}
\setcounter{figure}{0}

\renewcommand{\thesection}{\Alph{section}}

\pagebreak
\section*{Online Appendix A: Additional Methods and Results}

\renewcommand{\thesection}{A.\arabic{section}}

\section{Additional Methods}

\renewcommand{\theequation}{A\arabic{equation}}
% redefine the command that creates the equation number.
\renewcommand{\thetable}{A\arabic{table}}
\setcounter{equation}{0}  % reset counter 
\setcounter{figure}{0}
\setcounter{table}{0}

A suitable spatial discretization of equation (2) reads (\citealt{dornic05,giometto14}):
\begin{equation}
\begin{aligned}
\frac{d \rho_i}{d t}(t)= & \frac{1}{(\Delta x)^2} \left[ \rho_{i+1}(t)+\rho_{i-1}(t)-2\rho_i(t) \right] + r_i \rho_i(t)\left[1-\rho_i(t) \right] + \frac{\sigma}{\sqrt{\Delta x}} \sqrt{\rho_i(t)} \eta_i(t),
\label{FK_env1_discrete}
\end{aligned}
\end{equation}
where $i$ identifies the lattice site, the term $\sqrt{\Delta x}$ ensures proper normalization in the continuum limit (\citealt{doering05}) and $r_i=\delta_{I_i,1}$ depends on the local value of the resource profile $I$ (here, $\delta$ is the Kronecker's delta). The split-step method proposed in Dornic et al. (2005) was used to solve equation (\ref{FK_env1_discrete}). The spatial step in the numerical integration of equation (\ref{FK_env1_discrete}) was set to $\Delta x=0.5$, while the temporal step was chosen equal to $\Delta t=0.1$. The Courant-Friedrichs-Lewy condition for the diffusion equation $2 \Delta t / \Delta x^2 < 1$ was thus satisfied and $\Delta t / \Delta x < 1$. The numerical integration of equation (\ref{FK_env1_discrete}) with $\sigma=0$ was performed using the same numerical scheme, modified in the choice of $\rho^*$ (we refer the reader to  \citealt{dornic05} for notation and symbols), which in the deterministic case is $\rho^*=\alpha/\beta \left( e^{\beta \Delta t}-1\right) + \rho e^{\beta \Delta t}$. The deterministic equation was integrated with three choices of the growth rate $r$ in unfavorable regions of the landscape (where $I=0$), specifically $r=0$, $r=-0.01$ and $r=-0.1$. None of these choices for $r$ produced a slowdown of the front at large resource autocorrelation lengths, compared to small ones.

The spatial discretization of equation (3) reads:
\begin{equation}
\begin{aligned}
\frac{d \rho_i}{d t}(t)= & \frac{D}{(\Delta x)^2} \left[ \rho_{i+1}(t)+\rho_{i-1}(t)-2\rho_i(t) \right] - \frac{1}{2 \Delta x} \left[ g_{i+1} \rho_{i+1}(t) - g_{i-1} \rho_{i-1}(t) \right] \\  & + r_i \rho_i(t)\left[1-\frac{\rho_i(t)}{K} \right] + \frac{\sigma}{\sqrt{\Delta x}} \sqrt{\rho_i(t)} \eta_i(t),
\label{FK_env_discrete}
\end{aligned}
\end{equation}
where $g={d\phi}/{dx}[I(x)]$. The split-step method proposed in Dornic et al. (2005) was modified to solve equation (\ref{FK_env_discrete}), which contains an advection term that might cause an artificial loss of mass if the step sizes are too coarse. Such issue does not occur with the step sizes $\Delta x=0.6$ cm and $\Delta t=0.5$ min$^{-1}$ chosen here. The Courant-Friedrichs-Lewy condition for the diffusion equation $2D \Delta t / \Delta x^2 < 1$ was satisfied and $\Delta t / \Delta x < 1$. 

\section{Additional Results}

\subsection{Mean front propagation speed in heterogeneous landscapes}
Here we derive an approximation to the front propagation speed in the model equation (1), valid for large autocorrelation lengths and $\sigma$. We divide equation (1) by $K$ and $r_0$ and rescale time as $t' = r_0t$, which gives:
\begin{equation}
\frac{\partial \rho'(x,t')}{\partial t'} = \frac{D}{r_0} \frac{\partial^2 \rho'(x,t')}{\partial x^2} + \chi_I(x) \rho'(x,t') \left[ 1-\rho'(x,t') \right] +\frac{\sigma'}{\sqrt{r_0}}\sqrt{\rho'(x,t')} \ \eta(x,t'),
\end{equation}
where $\rho'=\rho/K$, $\sigma'=\sigma/\sqrt{K}$ and $\chi_I$ is the indicator function of the set of $x$ for which $I(x)>0$. We can further rescale space as $x' = \sqrt{\frac{D}{r_0}}x$ and rewrite equation (1) as:
\begin{equation}
\frac{\partial \rho'(x',t')}{\partial t'} = \frac{\partial^2 \rho'(x',t')}{\partial x'^2} + \chi_I(x') \rho'(x',t') \left[ 1-\rho'(x',t') \right] +\sigma'' \sqrt{\rho'(x',t')} \ \eta(x',t'),
\label{adimenseqA}
\end{equation}
where $\sigma''=\frac{\sigma' }{(rD)^{1/4}}$. In the following we will study the front propagation speed in the rescaled equation (\ref{adimenseq}), where we drop primes for convenience; one can recover the original dimensions by multiplying $t$ by $r_0$ and $x$ by $\sqrt{r_0/D}$.

The rationale for our approximation of the mean front propagation speed is as follows. Let $L$ be the finite length of a landscape and $T$ the time taken by the population to reach the end of such landscape ($x=L$), starting from a localized initial condition at $x=0$. For large values of autocorrelation length $c_L$ and large enough $\sigma$, due to the local extinctions caused by demographic stochasticity, most of the time $T$ is spent by the population trying to cross long patches of the landscape where $r=I=0$. We can therefore approximate the mean front propagation speed for large $c_L$ by computing the mean time that the front takes to cross an unfavorable patch of finite length $z$. Of course, such approximation is only valid when the waiting times dominate over the typical time scale of front propagation in favorable regions of the landscape. Therefore, the approximation can only hold for large enough values of the strength of demographic stochasticity $\sigma$. \looseness=-1

\subsubsection{Propagation past a patch of unfavorable landscape}
\begin{figure}[h]
\centering
\includegraphics{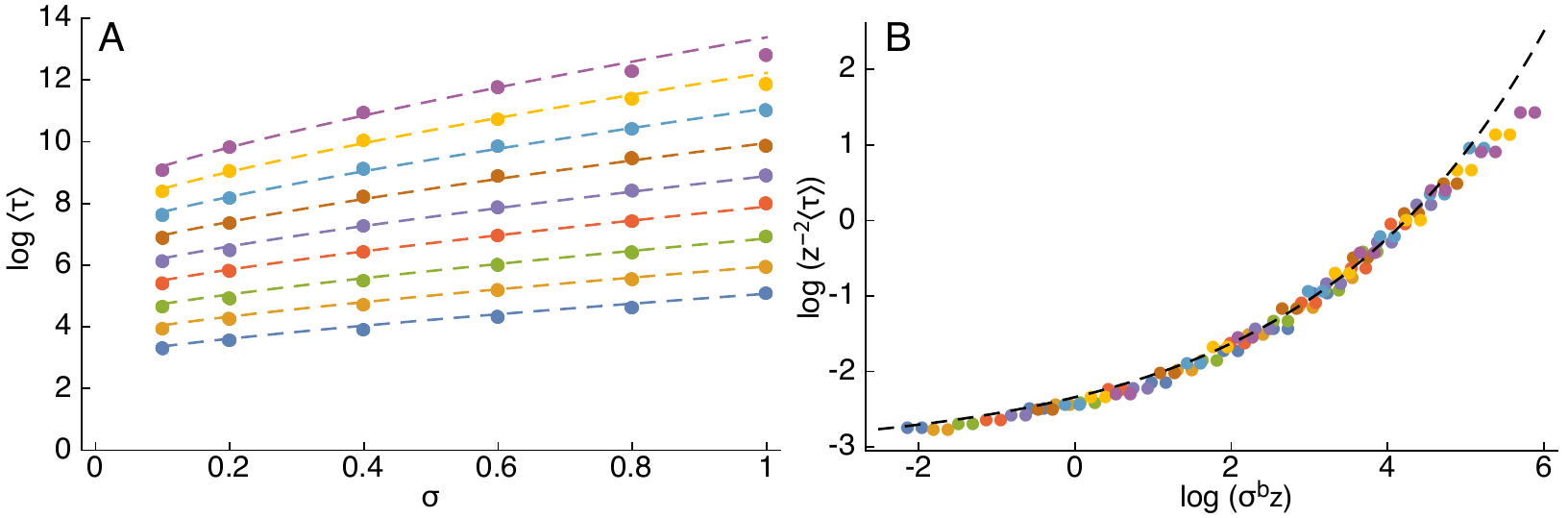}
\caption{Mean time $\langle \tau \rangle$ taken by a diffusing population subject to demographic stochasticity to cross patches of length $z$, calculated for different values of $z$ and $\sigma$ across $192$ integrations of equation (2). \textit{(A)} $\tau$ is a monotonically increasing function of $z$ and $\sigma$. Dots of identical color were computed with identical $z=21,\ 29,\ 49,\ 57,\ 79,\ 111,\ 156$ and $218$, from bottom to top. Lines are computed via equation (\ref{tauE}), the color code identifies the value of $z$ as for the dots. \textit{(B)} Simulation data collapse onto the same curve when $z^{-2} \tau$ is plotted against $\sigma^b z$, proving the assumption made in equation (\ref{tau}). Dots are color-coded as in panel (\textit A), the dashed black line shows the function $\mathbf{F}$ computed according to equation (\ref{tauE}).}
\label{tauF}
\end{figure}

\begin{figure}[H]
\centering
\includegraphics{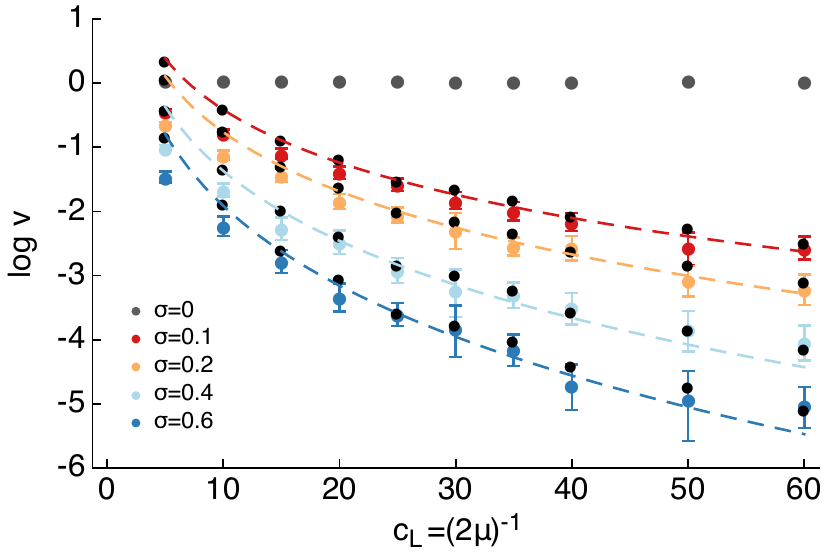}
\caption{The mean front speed $v$ decreases with increasing resource autocorrelation length $c_L=(2\mu)^{-1}$ ($\mu$ is the rate of the telegraph process used to generate the heterogeneous landscapes) and can be approximated by equation (\ref{vapprox}) for large $c_L$ (dashed lines). Colored data points highlight the mean speed $v$ computed by numerically integrating equation (2) and by fitting the mean front position versus time to a straight line. Different colors refer to different values of $\sigma$ according to the legend. Error bars display the $95$\% confidence interval for $v$, computed with $2 \cdot 10^3$ bootstrap samples. Error bars for $\sigma=0$ are smaller than symbols. Dashed lines are the mean front speed computed according to equation (\ref{vapprox}). Black dots are the approximation to the mean front speed computed as $v=L/T=L/\sum_{i \in Z} \langle \tau \rangle(z_i,\sigma)$, where $Z$ is the set of unfavorable windows in the numerical landscapes. Dashed lines and black dots may differ because the numerical landscapes were finite, thus the distribution of unfavorable window lengths may differ slightly from the exponential pdf with typical length $1/\mu=2 c_L$.}
\label{speed_fig}
\end{figure}

We computed numerically the mean time $\langle \tau \rangle$ taken by the front to cross a region of landscape where $I=0$, for different spatial extents of such region and different values of $\sigma$. We integrated numerically equation (2) in landscapes with resource profile
$I(x)=\theta(x-z)$, where $\theta$ is the Heaviside step function. Such landscapes consist of a resource profile $I(x)=r(x)=1$, except for $x \in (0,1]$, that is a finite patch of spatial extent $z$ at the left end of the landscape, where $I(x)=0$. The initial condition was $\rho(x,0)=0$ for $x>0$ and $\rho(0,0)=k$, where $k$ is the mean population density computed numerically by integrating equation (2) in a landscape of spatial extent $L=100$ with growth rate profile $r(x)=1$ for all $x\in[0,L]$. We fixed the Dirichlet boundary condition $\rho(0,t)=k$ and reflecting boundary conditions in $x=L$. We computed the mean time taken by the front to cross such unfavorable patch by measuring the first occurrence of $\rho(z)>10^{-3}k$ in time.
Fig. \ref{tauF}\textit A shows the mean time $\langle \tau \rangle$ taken by the population to cross unfavorable patches of various extents $z$, computed for various values of $\sigma$. Such mean time $\langle \tau \rangle$ is a monotonically increasing function of both $z$ and $\sigma$. To characterize the functional dependence of $\langle \tau \rangle(z,\sigma)$ on $z$ and $\sigma$, we note that in the limit $\sigma=0$ the dependence of $\langle \tau \rangle$ on $z$ is that of the deterministic diffusion equation with boundary condition $\rho(0,t)=1$, that is, $\tau(z,0) = C z^2$, where $C$ is the solution of $\mbox{erfc}\left( \frac 1{2\sqrt{C}} \right)=10^{-3}$, where $\mbox{erfc}$ is the complementary error function. We assume that $\langle \tau\rangle(z,\sigma) $ depends on $z$ and $\sigma$ through the functional form:
\begin{equation}
\langle \tau\rangle(z,\sigma) =C z^2 \mathbf{F}(z \sigma^b),
\label{tau}
\end{equation}
where $\mathbf F(x)$ is a function that goes to the constant $1$ for $x\to0$. We can verify the validity of equation (\ref{tau}) by plotting $z^{-2} \tau$ versus $z \sigma^b$ and varying $b$. Because we are able to find a value of $b=b^*$ for which data from the numerical integrations collapse onto one single curve (fig. \ref{tauF}\textit B), the assumption on the functional form of $\langle \tau \rangle$ is verified. To further identify the functional dependence of $\langle \tau\rangle$ on $z$ we plotted $\log [ \log (z^{-2} \langle \tau \rangle) - \log C ]$ vs $\log (z \sigma^{b})$ and observed that simulation data aligned along a straight line. Therefore, our numerical analysis suggests that the functional dependence of $\tau$ on $z$ and $\sigma$ is given by:
\begin{equation}
\langle \tau\rangle(z,\sigma) = C z^2 e^{ d\left( z \sigma^b \right)^a}.
\label{tauE}
\end{equation}

We estimated $b$ by maximizing the $R^2$ (coefficient of determination) of the least-squares linear fit of  $\log [ \log (z^{-2} \langle \tau\rangle) - \log C ]$ versus $\log (z \sigma^{b})$. The slope and intercept of the linear fit with maximum $R^2$ gave the estimate of $a$ and $d$. Fig. \ref{tauF} shows that equation (\ref{tauE}) reproduces the numerical data satisfactorily with the parameters $d=0.74$, $a=0.34$ and $b=2.25$, identified as outlined above.

\subsubsection{Approximation for the mean front propagation speed in heterogeneous landscapes}
For large values of the autocorrelation length $c_L=1/(2\mu)$ ($\mu$ is the rate of the telegraph process used to generate the heterogeneous landscapes, see Methods), most of the time taken by the front to propagate through a landscape of length $L$ is spent trying to cross finite stretches of the landscape where $r=I=0$. We can therefore approximate the front propagation speed as $v=L/T=L/\sum_{i=1}^N \langle \tau \rangle(z_i,\sigma)$ (black dots in fig. \ref{speed}), where $N$ is the number of unfavorable patches in $x\in[0,L]$ (of extent $z_i$) and $\langle \tau \rangle$ is the mean time taken to cross a patch of spatial extent $z_i$, estimated via equation (\ref{tauE}). In landscapes where unfavorable patches of length $z$ occur with probability $\mu e^{-\mu z}dz$, one can approximate the mean front propagation speed for large autocorrelation length $c_L$ as:
\begin{equation}
v=\frac{L}{ \frac{\mu L}{2} \int_0^L dz \langle \tau \rangle(z,\sigma) \mu e^{-\mu z} } = \frac{c_L^2}{{2\int_0^L dz \langle \tau \rangle(z,\sigma) e^{-z/(2c_L)}}},
\label{vapprox}
\end{equation}
where $\langle \tau \rangle(z,\sigma)$ is given by equation (\ref{tauE}) and $\mu L/2$ at the denominator is the average number of unfavorable patches in the landscape. If $L$ is comparable to $c_L$, one can substitute $\mu L/2$ with a more precise estimate, which is given in the next section. Fig. \ref{speed} shows that equation (\ref{vapprox}) gives a good approximation to the front propagation velocities computed in the numerical integrations, for large values of $c_L$.

\pagebreak 

\subsubsection{Correction to the average number of patches if $L$ is comparable to $c_L$}
We provide here a correction to the term $\mu L/2$ at the denominator of equation (\ref{vapprox}), which is relevant when $L\simeq c_L$. If the first patch at $x=0$ is favorable (i.e., $r>0$), the average number of unfavorable patches in a landscape of length $L$ can be computed as follows. Let $z_i$ be the rightmost coordinate of each patch in the landscape. The average number of unfavorable patches is equal to:
\begin{equation*}
\langle N \rangle =  \sum_{n=1}^\infty n \mathbb{P}\left[z_{2n}<L \cap z_{2n+1} \geq L\right] + \sum_{n=1}^\infty n \mathbb{P}[z_{2n-1}<2 \cap z_{2n} \geq L].
\end{equation*}
Using properties of the exponential distribution of patch lengths one has:
\begin{align*}
\mathbb{P}&\left[z_{2n}<L \cap z_{2n+1} \geq L\right]\\&=\mu^{2n+1}\int_0^L dz_1 e^{-\mu z_1}\int_{z_1}^L  dz_2 e^{-\mu(z_2-z_1)}\cdots \int_{z_{2n-1}}^L dz_{2n}e^{-\mu(z_{2n}-z_{2n-1})}\int_{L}^\infty dz_{2n+1}e^{-\mu(z_{2n+1}-z_{2n})}\\&=\frac{e^{-\mu L}}{(2n)!}(\mu L)^{2n},\\
\mathbb{P}&[z_{2n-1}<2 \cap z_{2n} \geq L]\\&=\mu^{2n}\int_0^L dz_1 e^{-\mu z_1}\int_{z_1}^L  dz_2 e^{-\mu(z_2-z_1)}\cdots \int_{z_{2n-2}}^L dz_{2n-1}e^{-\mu(z_{2n-1}-z_{2n-2})}\int_L^\infty dz_{2n}e^{-\mu(z_{2n}-z_{2n-1})}\\&=\frac{e^{-\mu L}}{(2n-1)!}(\mu L)^{2n-1}
\end{align*}
and therefore:
\begin{equation*}
\langle N \rangle = \sum_{n=1}^\infty n \left[ \frac{e^{-\mu L}}{(2n)!}\left( \mu L \right)^{2n}+ \frac{e^{-\mu L}}{(2n-1)!}(\mu L)^{2n-1} \right]=\frac{\mu L}{2}+\frac{e^{-\mu L}}{2} \sinh(\mu L),
\end{equation*}
where $\sinh$ is the hyperbolic sine function. One can repeat the same analysis in the case where the first patch at $x=0$ is unfavorable (i.e., $r=0$). In this case one finds:
\begin{equation*}
\langle N \rangle = \sum_{n=1}^\infty n \left[ \frac{e^{-\mu L}}{(2n-2)!}\left( \mu L \right)^{2n-2}+ \frac{e^{-\mu L}}{(2n-1)!}(\mu L)^{2n-1} \right]=\frac{\mu L}{2}+\frac34 + e^{-\mu L}.
\end{equation*}
Finally, if the first patch is favorable or unfavorable with equal probabilities, then:
\begin{equation*}
\langle N \rangle = \frac12 \left[ \frac{\mu L}{2}+\frac{e^{-\mu L}}{2} \sinh(\mu L) \right] + \frac12 \left[ \frac{\mu L}{2}+\frac34 + e^{-\mu L} \right] = \frac12 + \frac{\mu L}2.
\end{equation*}
If $L \gg \frac 2\mu = 4 c_L$, the average number of unfavorable patches in a landscape of length $L$ tends to $\frac{\mu L}2$.\looseness=-1

\pagebreak
\subsection{Fluctuations of the invasion time}
\subsubsection{Fluctuations of the time taken to cross a patch of unfavorable landscape}
\begin{figure}[H]
\centering
\includegraphics{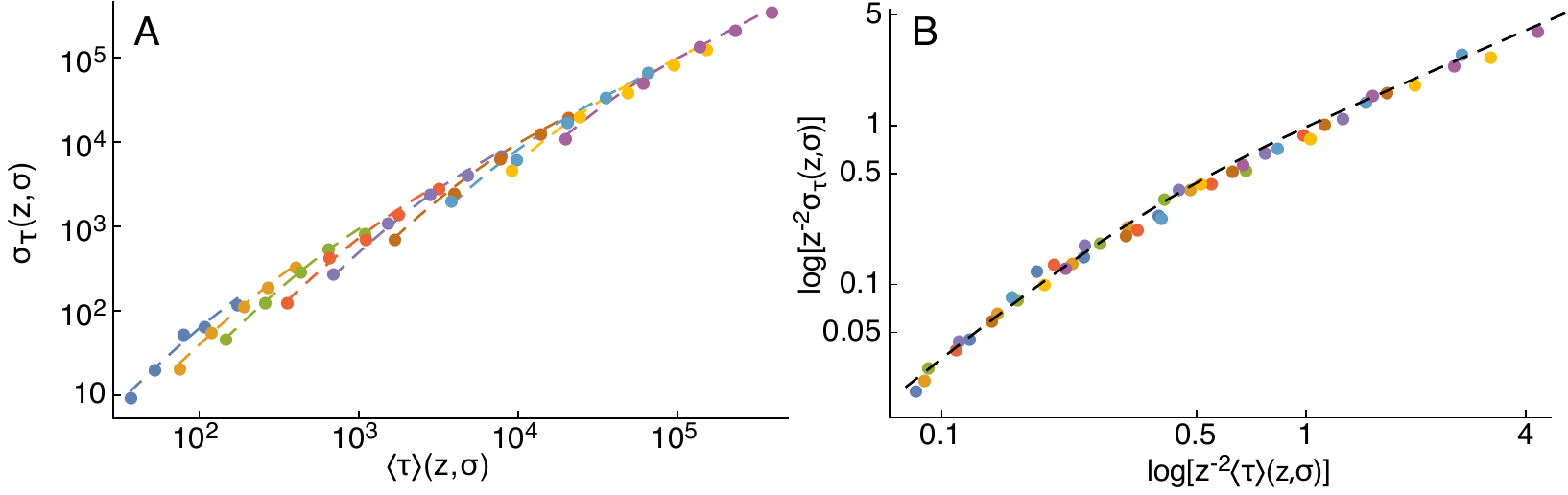}
\caption{\textit{(A)} Standard deviation $\sigma_\tau$ of the time taken by a diffusing population subject to demographic stochasticity to cross patches of length $z$, calculated for different values of $z$ and $\sigma$ across $96$ integrations of equation (\ref{adimenseq}) (double logarithmic plot). Different colors refer to different values of $\sigma$, from $\sigma=0.1$ (blue dots at the bottom left corner) to $\sigma=1.4$ (violet dots at the top right corner). Dashed lines are computed with equation (\ref{sigmafunc}). \textit{(B)} Simulation data collapse onto the same curve when $z^{-2} \sigma_\tau$ is plotted against $z^{-2} \langle \tau \rangle$, proving the validity of equation (\ref{sigmascaling}). Dots are color-coded as in panel (\textit A), the dashed black line shows the function $\mathbf{S}$ computed according to equation (\ref{sigmafunc}).}
\label{stdtau}
\end{figure}

In this section we study the fluctuations of the total invasion time in heterogeneous landscapes of finite size $L$. To this end, we first characterize the standard deviation $\sigma_\tau$ of the time $\tau$ taken by a diffusing population subject to demographic stochasticity to cross an unfavorable patch ($r=0$) of spatial extent $z$. Inspection of the numerical results shows (fig. \ref{stdtau}\textit B) that $ z^{-2}\sigma_\tau$ is a function of $z^{-2} \langle \tau\rangle(z,\sigma) $, that is:
\begin{equation}
\sigma_\tau(z,\sigma) = z^2 \mathbf{S}\left( z^{-2} \langle \tau\rangle(z,\sigma)  \right),
\label{sigmascaling}
\end{equation}

where $\mathbf{S}(x)$ is a function that goes to $0$ for $x \to 0$. In fact, data from the numerical integrations of equation (2) in landscapes with resource profile $I(x)=\theta(x-z)$ ($\theta$ is the Heaviside step function, the same numerical data were used to derive equation \ref{tauE}) collapse on the same curve when $z^{-2} \sigma_\tau$ is plotted against $z^{-2} \langle \tau\rangle(z,\sigma) $ (fig. \ref{stdtau}\textit B). The functional form:
\begin{equation}
\sigma_\tau(z,\sigma)=\langle \tau\rangle(z,\sigma)  \left[1-e^{-k z^{-2} \langle \tau\rangle(z,\sigma) } \right]
\label{sigmafunc}
\end{equation}
is found to provide a good fit to the numerical data, with the best-fit estimate of the coefficient $k=4.17$ (dashed lines in fig. \ref{stdtau}).

\subsubsection{Fluctuations of the total invasion time in heterogeneous landscapes}
\begin{figure}[H]
\centering
\includegraphics{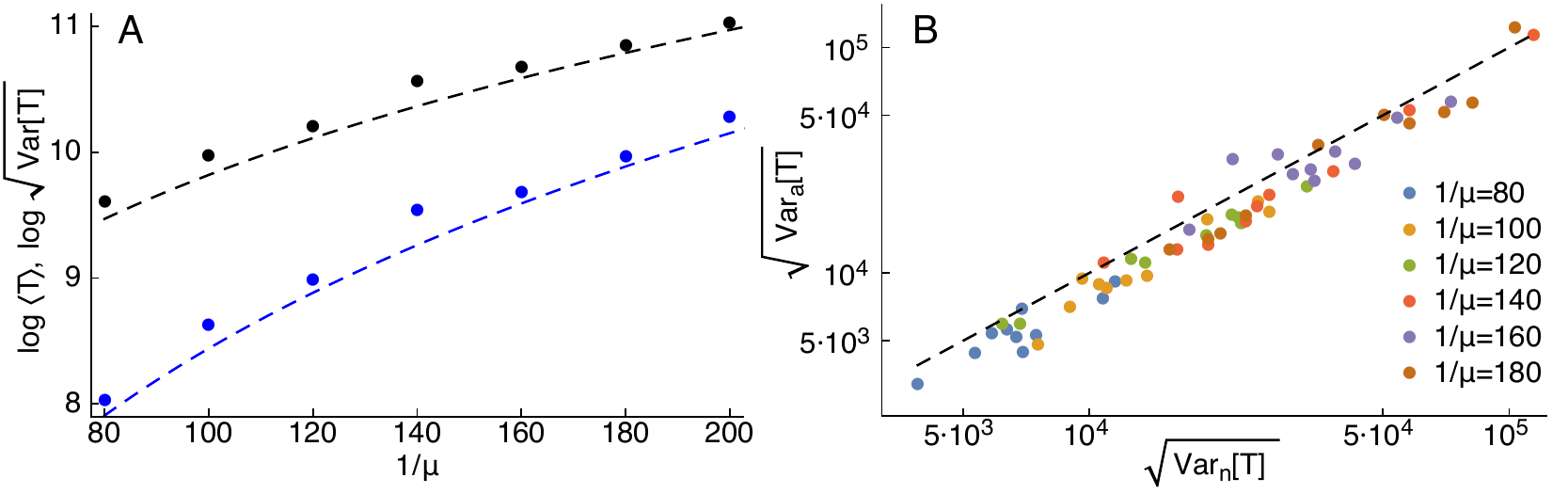}
\caption{\textit{(A)} Mean total time $\langle T \rangle$ (black dots) of invasion and its standard deviation (blue dots) in numerical integrations of equation (2) in square-wave landscapes of length $L=1400$, that is, landscapes composed of alternated favorable and unfavorable patches of length $1/\mu$ (means and standard deviations were computed across $200$ integrations for each value of $1/\mu$). The numerical estimates for $\langle T \rangle$ and $\sqrt{\mbox{Var}[T]}$ are well approximated by the approximations $\langle T \rangle = \frac{\mu L}{2} \tau(1/\mu,\sigma)$ (black dashed line) and by equation (\ref{varexact}) (blue dashed line). \textit{(B)} Numerically computed standard deviations $\sqrt{\mbox{Var}_n[T]}$  (double logarithmic plot) of the total time $T$ of invasion in numerical integrations of equation (2) in landscapes with exponentially-distributed favorable and unfavorable patches are well approximated by the theoretical approximation $\sqrt{\mbox{Var}_a[T]}$, computed according to equation (\ref{varexact}). Each dot represents one landscape of length $L=2000$ and mean patch length $1/\mu$ according to the legend. Such landscapes were generated with the same procedure outlined in the Methods. To compute $\sqrt{\mbox{Var}_n[T]}$, we performed $96$ numerical integrations for each landscape. The dashed black line is the 1:1 line. Numerical estimates and theoretical approximations are calculated with $\sigma=0.4$ in both panels.}
\label{variance_sq_exp}
\end{figure}

We can use equation (\ref{sigmafunc}) to approximate the variance of the total invasion time $T$ (i.e., the time after which the density $\rho(L,T)$ is larger than a threshold density value) in heterogeneous landscapes composed of favorable and unfavorable patches. In fact, the variance of the total invasion time in our simplified model, where we neglect the time spent by the front in propagating through favorable patches, and further assuming that the times spent to cross each unfavorable patch are independent from each other, is given by:
\begin{equation}
\mbox{Var}[T]=\sum_{i=1}^N \sigma_\tau^2[\langle \tau \rangle(z_i,\sigma)],
\label{varexact}
\end{equation}
where $N$ is the number of unfavorable patches in $x\in[0,L]$ (patches of extent $z_i$) and $\sigma_\tau$ is given by equation (\ref{sigmafunc}). We show in fig. \ref{variance_sq_exp} that equation (\ref{varexact}) gives a good estimate of the variance of the total invasion time in heterogeneous landscapes. Details are provided in the figure caption.

\pagebreak
\subsection{Front propagation at different mean resource densities}
\begin{figure}[H]
\centering
\includegraphics{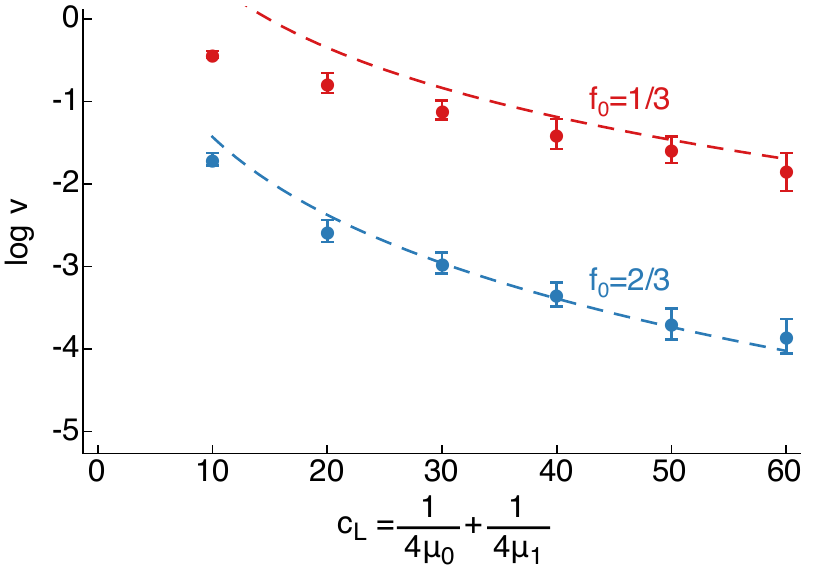}
\caption{Mean front propagation speed in landscapes with favorable and unfavorable habitats occurring at frequencies different from $f_1=f_0=1/2$. Red dots display the mean front speed in $96$ replicated invasions in different landscapes with frequency of unsuitable habitat $f_0=1/3$. favorable patches lengths were distributed exponentially with rate $\mu_1=3/(8 c_L)$ and unfavorable ones with rate $\mu_0=3/(4 c_L)$. Blue dots display the mean front speed in $96$ replicated invasions in different landscapes with frequency of unsuitable habitat $f_0=2/3$. Error bars display the $95$\% confidence interval for $v$, computed with $2 \cdot 10^3$ bootstrap samples. favorable patches lengths were distributed exponentially with rate $\mu_1=3/(4 c_L)$ and unfavorable ones with rate $\mu_0=3/(8 c_L)$. Dashed lines show mean front  speeds approximated via equation (5) of the main text.}
\label{mean1323}
\end{figure}

\begin{figure}[h]
\centering
\includegraphics{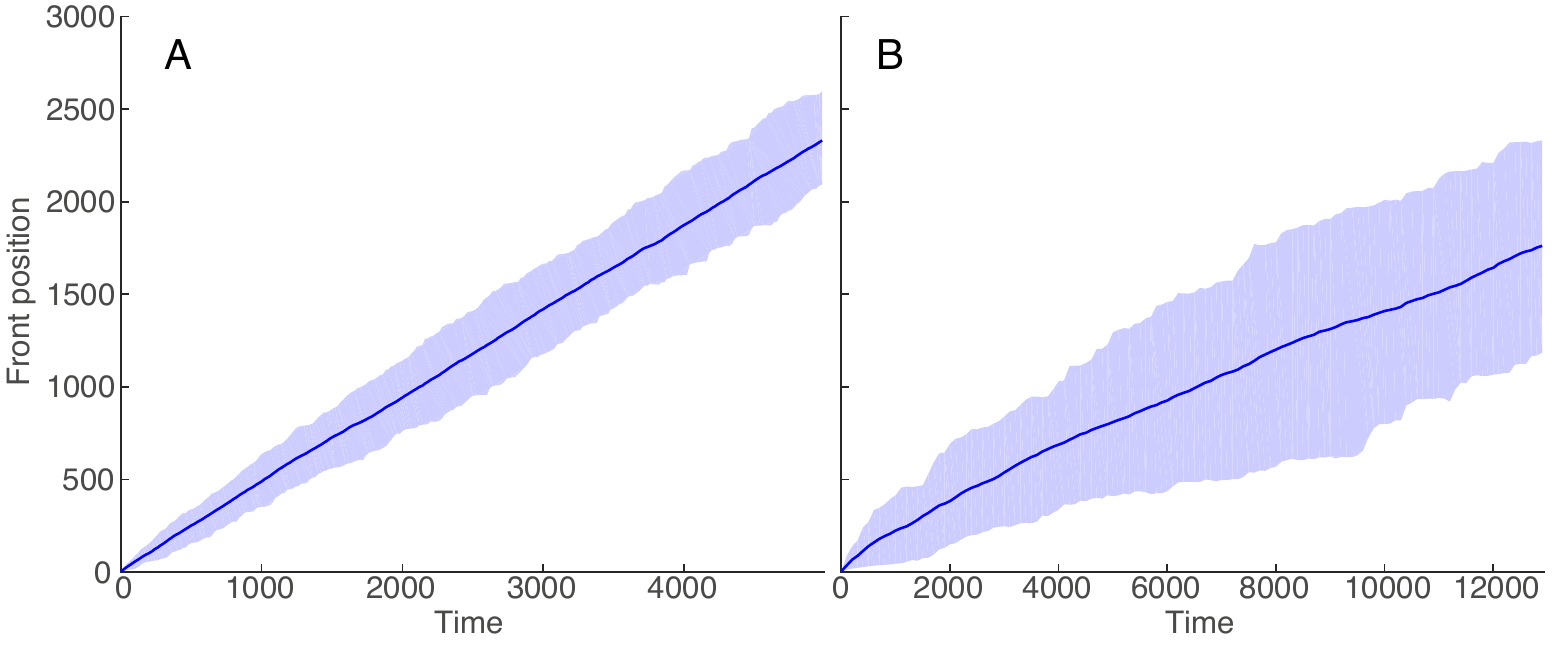}
\caption{Mean position of the front (blue lines) and $68$\% confidence interval (shaded regions) in numerical integrations of the model equation (2) with $\sigma=0.1$ and resource autocorrelation lengths $c_L=5$ (\textit A) and $c_L=20$ (\textit B).}
\label{mean_front}
\end{figure}

\begin{figure}[h]
\centering
\includegraphics{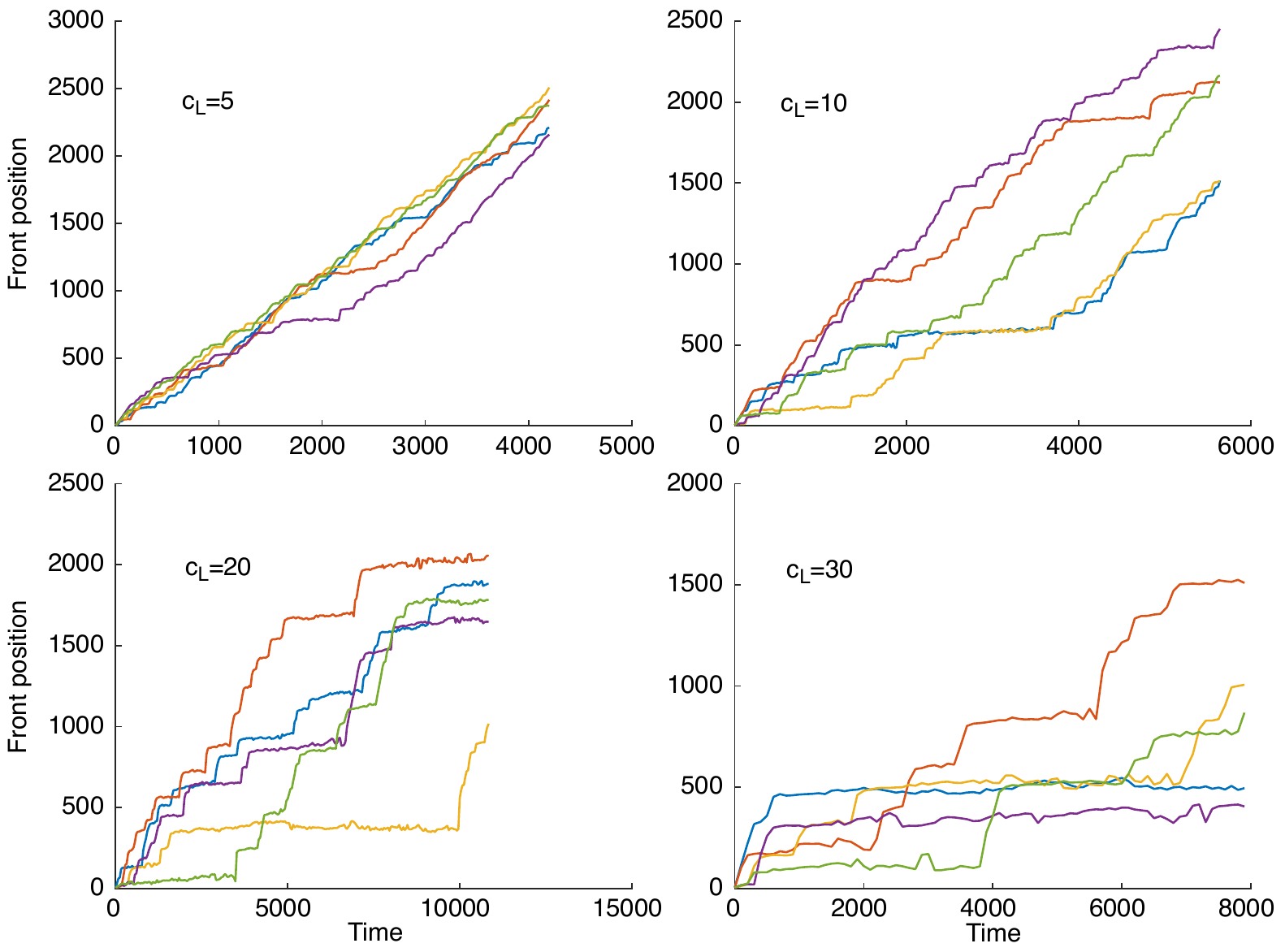}
\caption{Examples of front propagation in numerical integrations of the model (equation 2) in landscapes with different resource autocorrelation lengths $c_L$ and fixed amplitude of demographic stochasticity $\sigma=0.2$.}
\label{front_simulations}
\end{figure}

\begin{figure}
\centering
\includegraphics{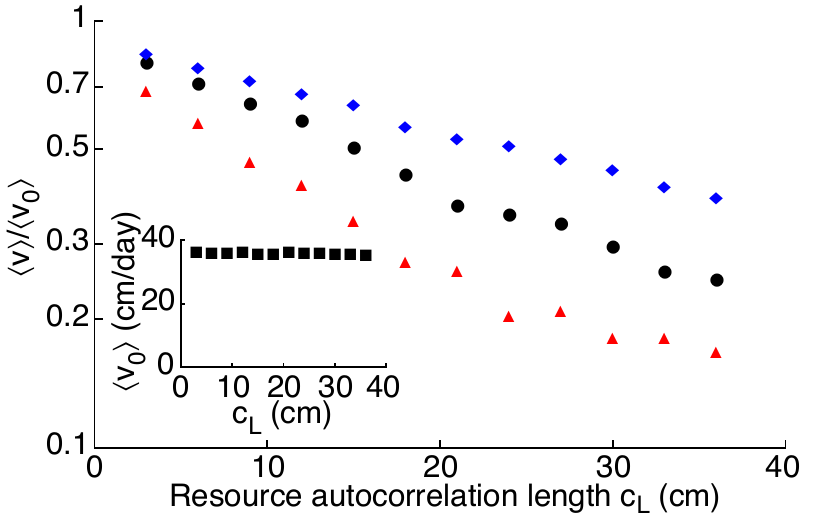}
\caption{Front propagation computed in numerical integrations of the model equation (3) (with spatial discretization equation \ref{FK_env_discrete}). The mean invasion speed decreases with increasing resource autocorrelation length $c_L=\Delta L/(2\lambda$) ($\lambda$ is the transition probability of the Markov Chain used to generate the heterogeneous landscapes and $\Delta L$ is the experimental distance between LEDs, see Methods) and is a decreasing function of the amplitude of demographic stochasticity $\sigma$ (log-linear plot; black dots: $\sigma=0.4$ min$^{-1/2}$; red triangles: $\sigma=0.7$ min$^{-1/2}$). The mean speed of invasion is larger in the absence of directed movement towards resources (blue diamonds computed with $\sigma=0.4$ min$^{-1/2}$ and $\phi=0$). Invasion speeds are reported here divided by the mean front speed $\langle v_0 \rangle$ at $\sigma=0$ min$^{-1/2}$, that is constant for different values of $c_L$ (inset). The mean front speed for each value of $c_L$ and $\sigma$ was calculated by integrating equation (3) along $150$ different landscapes with identical $c_L$ and fitting the mean front position versus time in the asymptotic propagation regime.}
\label{speed_photo}
\end{figure}

Other works (e.g., \citealt{dewhirst09}) have studied the propagation of invasion fronts in landscapes with different average amounts of resources. One may wonder whether the slowdown effect caused by varying resource autocorrelation lengths of the resource distribution might also be found in landscapes endowed with mean percentages of suitable habitat different from $f_1=f_0=1/2$. To show that such slowdown effect occurs also when the suitable and unsuitable habitats occur at different frequencies throughout the landscape, we have integrated equation (2) on landscapes endowed with various resource autocorrelation lengths and mean frequency of suitable (i.e., $r>0$) and unsuitable (i.e., $r=0$) habitat equal to $f_1=1/3$ and $f_0=2/3$, respectively. Such landscapes were generated as follows: we extracted the length of each favorable and unfavorable patch from exponential distributions with rate $\mu_1=3/(4 c_L)$ and $\mu_0 = 3/(8 c_L)$, respectively, so that the resource autocorrelation length was $c_L$ and the frequencies of favorable/unfavorable habitat were as desired. Additionally, we have integrated equation (2) on the same landscapes switching each favorable patch of the landscape with an unfavorable one, so that favorable habitats occurred with frequency $f_1=2/3$ (and thus unfavorable habitats with frequency $f_0=1/3$). Fig. \ref{mean1323} shows that increasing the mean frequency of suitable habitat increases the invasion speed, but the slowdown effect caused by varying resource autocorrelation lengths is also present when favorable and unfavorable habitats occur at frequencies different from $1/2$. Furthermore, equation (5) can be used to approximate the mean speed of invasion for large $c_L$ at values of $f_0$ different from $1/2$, as shown by the agreement between dashed lines and simulation data points in fig. \ref{mean1323}.

\clearpage
\section{Additional Tables}

\begin{table}[h]
\caption{Mixed-effect test statistics for all choices of density threshold $\bar \rho$}
\begin{center}
\begin{tabular}{|c|l|r|r|r|r|r|} \hline
Threshold $\bar \rho$	&	\multicolumn{2}{ r |}{Value}	&	Std. Error	&	df	&	$t$-value	&	$p$-value	\\ \hline
$45$ cm$^{-1}$	&	Intercept	&	$57.15$	&	$3.65$	&	$44$	&	$15.65$	&	$p<10^{-4}$	\\
	&	Autocorrelation length	&	$-11.31$	&	$4.94$	&	$9$	&	$-2.29$	&	$p=0.0480$	\\ \hline
$60$ cm$^{-1}$	&	Intercept	&	$45.98$	&	$3.27$	&	$44$	&	$14.04$	&	$p<10^{-4}$	\\
	&	Autocorrelation length	&	$-11.61$	&	$4.43$	&	$9$	&	$-2.62$	&	$p=0.0279$	\\ \hline
$75$ cm$^{-1}$	&	Intercept	&	$45.27$	&	$2.88$	&	$44$	&	$15.70$	&	$p<10^{-4}$	\\
	&	Autocorrelation length	&	$-9.65$	&	$3.90$	&	$9$	&	$-2.47$	&	$p=0.0355$	\\ \hline
$90$ cm$^{-1}$	&	Intercept	&	$36.65$	&	$2.84$	&	$44$	&	$12.91$	&	$p<10^{-4}$	\\
	&	Autocorrelation length	&	$-9.04$	&	$3.85$	&	$9$	&	$-2.35$	&	$p=0.0433$	\\ \hline
$105$ cm$^{-1}$	&	Intercept	&	$35.91$	&	$3.04$	&	$44$	&	$11.83$	&	$p<10^{-4}$	\\
	&	Autocorrelation length	&	$-10.79$	&	$4.11$	&	$9$	&	$-2.62$	&	$p=0.0276$	\\ \hline
\end{tabular}
\end{center}
\label{me_stats_all}
\bigskip{}
{\footnotesize Mixed-effect test statistics testing the speed of front propagation, with the autocorrelation length treatment as single fixed effect and time/replicate as random effect. The treatment with small autocorrelation length had $5$ replicates, the treatment with large autocorrelation length had $6$ replicates. Different lines refer to different threshold values $\bar \rho$ at which the front position was measured.}
\end{table}

\clearpage
\section{Additional Figures}


\begin{figure}
\centering
\includegraphics{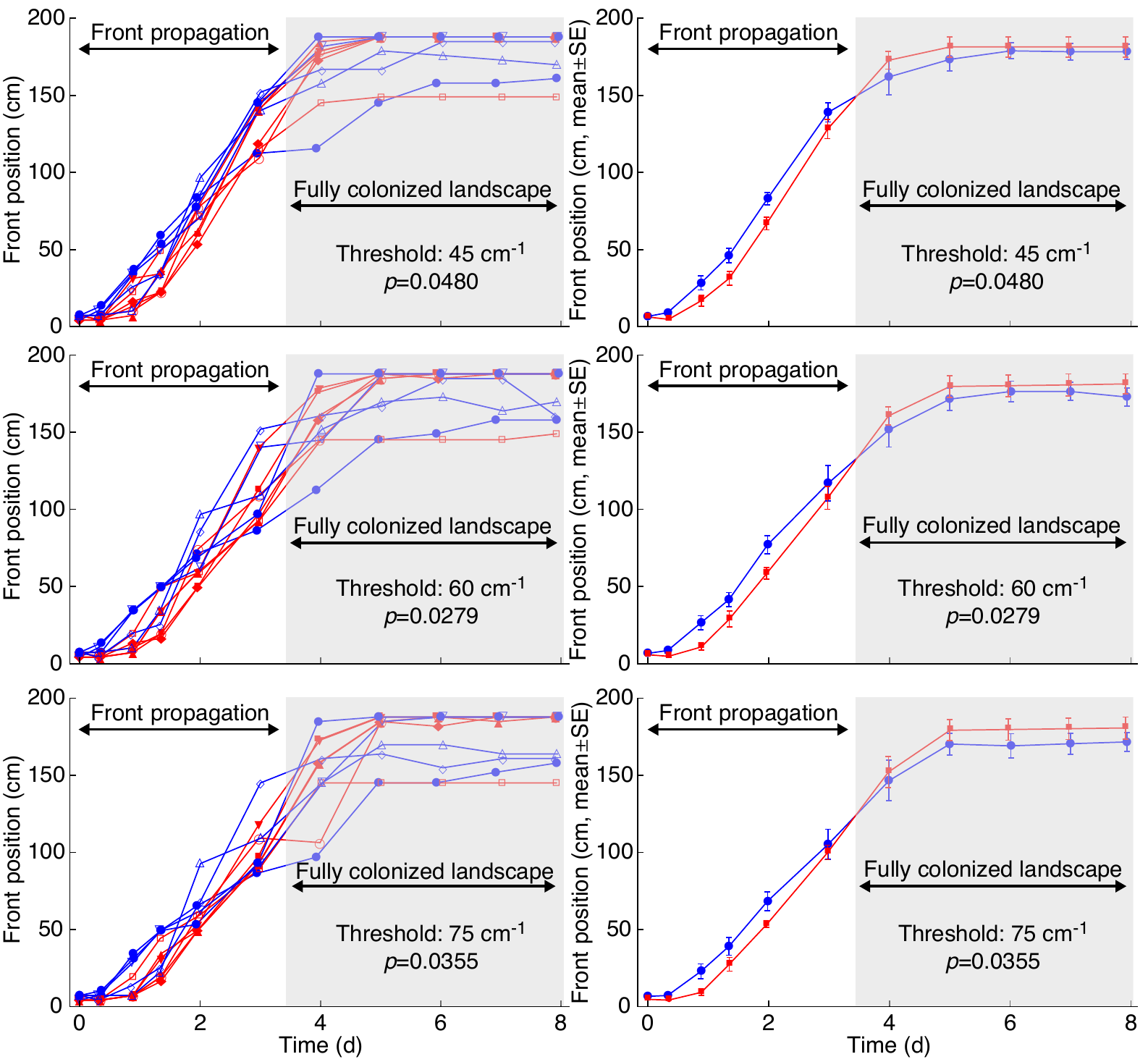}
\caption{Experimental spread in autocorrelated landscapes. Left: position of the front in each experimental replicate, identified by different symbols. Red and blue lines and symbols refer to replicates with identical large (red) or small (blue) resource autocorrelation length. Right: mean ($\pm$SE) position of the front, calculated among replicates with identical large (red) or small (blue) resource autocorrelation length. Different rows refer to different threshold density values used to identify the position of the front. The gray shaded regions identify data points collected when at least one replicate had colonized the whole landscape. To avoid border effects, we excluded such points from the statistical analysis. In fact, at least one replicate with small autocorrelation length had reached the end of the landscape at time $t=4$ d, and might have spread even further in a longer landscape. The reported $p$-values show that the autocorrelation treatment had a significant effect on the front propagation regardless of the choice of density threshold.}
\label{experiment_all}
\end{figure}

\begin{figure}
\centering
\includegraphics{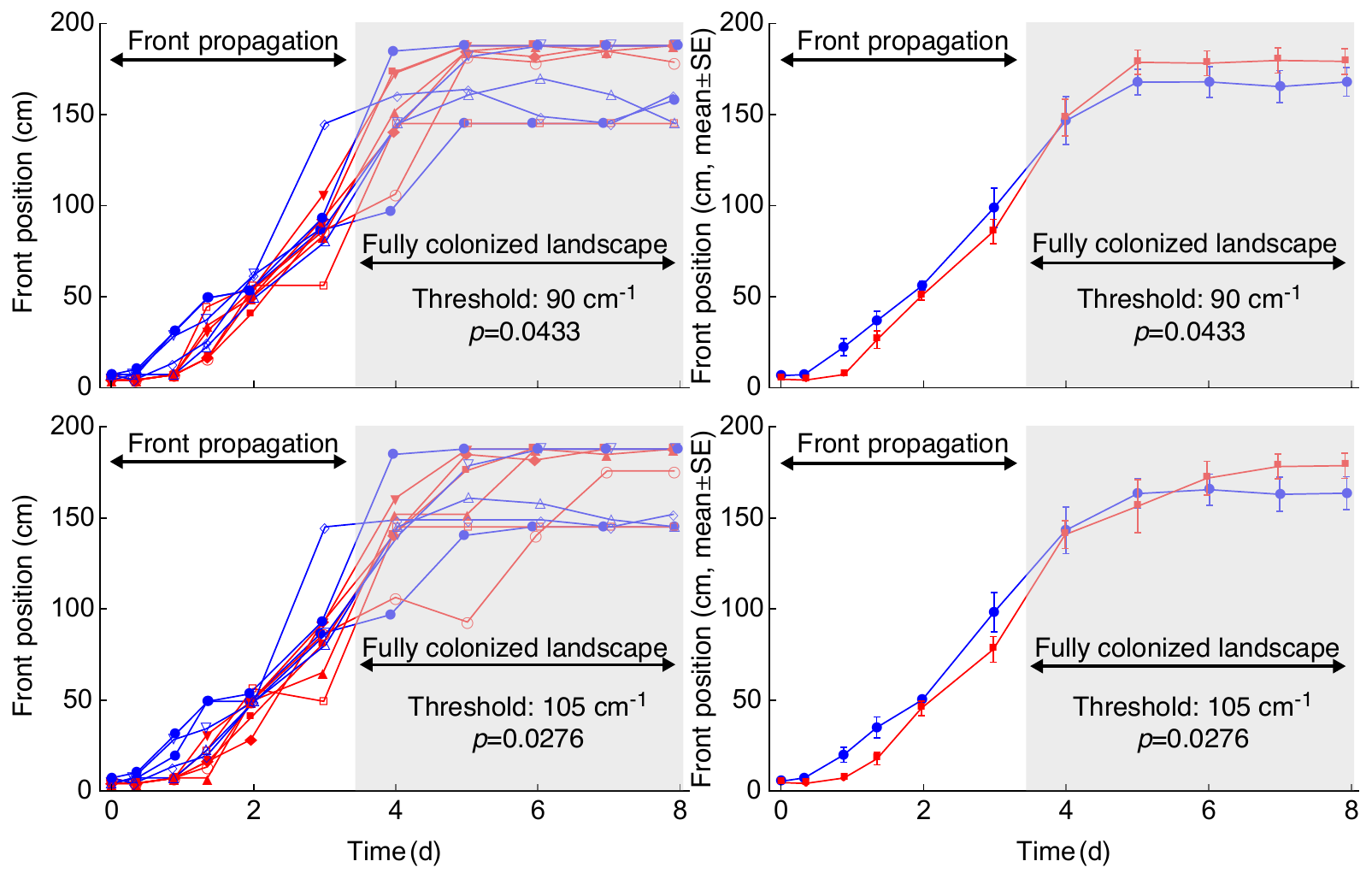}
\caption{Experimental spread in autocorrelated landscapes. Left: position of the front in each experimental replicate, identified by different symbols. Red and blue lines and symbols refer to replicates with identical large (red) or small (blue) resource autocorrelation length. Right: mean ($\pm$SE) position of the front, calculated among replicates with identical large (red) or small (blue) resource autocorrelation length. Different rows refer to different threshold density values used to identify the position of the front. The gray shaded regions identify data points collected when at least one replicate had colonized the whole landscape. To avoid border effects, we excluded such points from the statistical analysis. In fact, at least one replicate with small autocorrelation length had reached the end of the landscape at time $t=4$ d, and might have spread even further in a longer landscape. The reported $p$-values show that the autocorrelation treatment had a significant effect on the front propagation regardless of the choice of denstiy threshold.}
\label{experiment_all}
\end{figure}

\end{document}